\documentclass[fleqn,10pt]{wlscirep}
\usepackage[utf8]{inputenc}
\usepackage[T1]{fontenc}
\usepackage{eucal}
\usepackage{multirow}
\usepackage{tabularx}
\usepackage{caption}
\usepackage{rotating}
\usepackage{color}
\usepackage{float}
\usepackage{lmodern}
\usepackage{bm}
\usepackage{amssymb}

\usepackage{wasysym}
\usepackage{amsthm}
\newtheorem{theorem}{Theorem}[section]

\newtheorem{Proposition}{Proposition}[section]

\usepackage{soul}

\newcommand{\rv}[1]{{\color{black}#1}}

\title{On Fast Simulation of Dynamical System with Neural Vector Enhanced Numerical Solver 
}

\author[1,*]{Zhongzhan Huang}
\author[2,*]{Senwei Liang}
\author[3]{Hong Zhang}
\author[4]{Haizhao Yang}
\author[1,$^\dag$]{Liang Lin}
\affil[1]{School of Computer Science and Engineering, Sun Yat-sen University,  Guangzhou, China}
\affil[2]{Department of Mathematics, Purdue University, West Lafayette, IN, USA}
\affil[3]{Mathematics and Computer Science Division, Argonne National Laboratory,  Lemont, IL, USA}
\affil[4]{Department of Mathematics, University of Maryland College Park, College Park, MD, USA}
\affil[$^\dag$]{Correspondence should be addressed to: linliang@ieee.org}

\affil[*]{these authors contributed equally to this work, are listed with alphabetical order}

\begin{abstract}

The large-scale simulation of dynamical systems is critical in numerous scientific and engineering disciplines. However, traditional numerical solvers are limited by the choice of step sizes when estimating integration, resulting in a trade-off between accuracy and computational efficiency. To address this challenge, we introduce a deep learning-based corrector called Neural Vector (NeurVec), which can compensate for integration errors and enable larger time step sizes in simulations. Our extensive experiments on a variety of complex dynamical system benchmarks demonstrate that NeurVec exhibits remarkable generalization capability on a continuous phase space, even when trained using limited and discrete data. NeurVec significantly accelerates traditional solvers, achieving speeds tens to hundreds of times faster while maintaining high levels of accuracy and stability. Moreover, NeurVec's simple-yet-effective design, combined with its ease of implementation, has the potential to establish a new paradigm for fast-solving differential equations based on deep learning.

\end{abstract}
\begin{document}

\flushbottom
\maketitle
%
%
\thispagestyle{empty}

\section*{Introduction}
\begin{figure*}[t]
    \centering
    \includegraphics[width=0.9\linewidth]{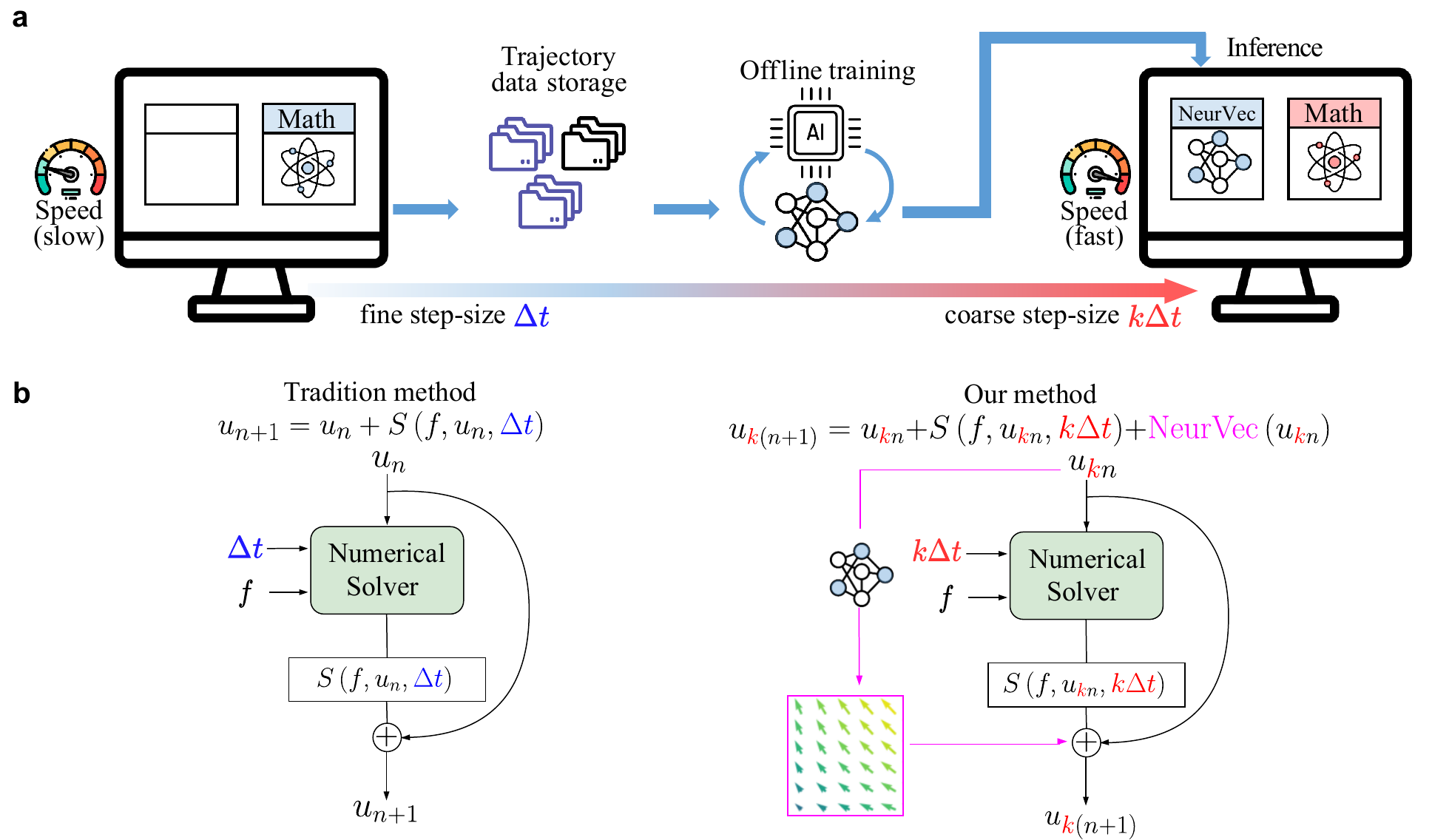}
    \caption{\textbf{The structure of NeurVec.} \textbf{a}, \rv{Deployment of NeurVec for fixed step size solver}. During offline training, NeurVec learns from the solutions of high accuracy to characterize the error caused by the use of a coarse step size. During inference, NeurVec applies to the solver and integrates with the coarse step size. \textbf{b}, Comparison of the tradition solver and solver armed by NeurVec. Left: Forward loop of traditional solver with step size $\Delta t$. Right: Forward loop of NeurVec with step size $k\Delta t$. }
    \label{fig:1}
\end{figure*}

Dynamical systems are widely used to characterize the time dependence of the physical states and to model phenomena that change with time.~\cite{bottcher2022ai,stuart1998dynamical,harlim2021machine} Studying the temporal evolution of dynamical systems and their statistics can help uncover the physics behind the dynamics and predict future states of the systems.~\cite{harlim2021machine} Typically, a time-dependent $d$-dimensional state $u(t)$ is depicted by a system of ordinary differential equations (ODEs): 
\begin{align}
\frac{\text{d}\mathbf{u}(t)}{\text{d}t} = \mathbf{f}[\mathbf{u}(t)], \quad \mathbf{u}(0) = \mathbf{c}_0,
\label{eqn:ode}
\end{align}
where $\mathbf{c}_0$ represents an initial condition. This system arises in many science and engineering fields such as climate change,\cite{kou2019nitrogen,benn2019general} air pollution,\cite{owoyele2022chemnode,zhang2011chemical}stable financial systems,\cite{delpini2013evolution} power grid management,\cite{gholami2022impact,schafer2018dynamically} transportation networks,\cite{saberi2020simple,fan2020network} and medical analysis and drug discovery.\cite{aulin2021design,butner2021mathematical,wicha2017general}
 To obtain a numerical solution of \eqref{eqn:ode}, one may employ an integration method~\cite{butcher2016numerical,ames2014numerical}  (Fig.~\ref{fig:1}b) given by the iterative formula 
 \begin{align}
     \mathbf{u}_{n+1} = \mathbf{u}_{n} + S(\mathbf{f}, \mathbf{u}_{n}, \Delta t_n), \quad \mathbf{u}_0 = \mathbf{c}_0, \quad n=0,1,\cdots,
     \label{eq:iterative}
 \end{align}
 where $S$ represents a numerical scheme (for example, $S(f, \mathbf{u}_{n}, \Delta t_n):=f(\mathbf{u}_{n})\Delta t_n$ when the Euler method\cite{shampine2018numerical} is used), $\Delta t_n$ is the step size of the $n$th time step, and $\mathbf{u}_{n}\in \mathbb{R}^d$ is an approximated solution at time $\sum_{i=0}^n \Delta t_n$. 
 When approximating a solution at a specific time given an initial condition, we readily customize accuracy and speed via tuning integration strategies (e.g., different scheme $S$ and step size $\Delta t_n$ selection). 
 However, many real-world applications\cite{figueiras2019qmblender,xi2019survey,zhang2020cumuliform,scher2021ensemble} require simulating many trajectories. In particular, large-scale simulation, which produces forecasts on a set of initial conditions simultaneously, is more useful for these applications. Compared with a single simulation, ensemble-based large-scale simulation is a computationally challenging problem but plays a critical role in a variety of demanding applications. For illustration, we present a few scenarios of such simulation.

\noindent\textbf{Fast simulation.} Sine late 2019, the epidemic of COVID-19 has raged around the world, hitting the global health and economy.\cite{hu2021characteristics} Scientists need to perform simulations of virus propagation under thousands of different circumstances.\cite{bousquet2022deep,beira2021differential} These predictions provide the scientific reference for governments to make quick responses and control policies.\cite{choi2021optimal,hsiang2020effect} The virus, such as the Delta and Omicron variants, spreads rapidly and mutates frequently.\cite{tregoning2021progress} A slow simulation may lead to a delay in decision-making and worsen the situation. 

\noindent\textbf{Synchronous simulation.} Particle systems are a graphical technique that simulates complex physical effects (such as smoke,\cite{yuan2019wave} water flow,\cite{tumanov2021data} and object collision.\cite{kolb2004hardware}). This is widely used in  applications in games, movies, and animation.\cite{luo2022using} These applications involve synchronously simulating thousands of particles at one time. Yet  supporting the real-time simulation of these particle motions with satisfactory visual enjoyment is expensive. 

\noindent\textbf{Reliable model.} Weather forecasting is beneficial for making a proper plan for production and living.\cite{bellprat2019towards,touma2021human} A single forecast of the weather model essentially suffers from considerable errors introduced by the highly simplified model formulation and the chaotic nature of the atmosphere evolution equations. In order to avoid a misleading single forecast, ensemble forecasting\cite{palmer2013singular,wu2021ensemble,popov2021multifidelity} presents a range of possible future weather states through conducting simulations from multiple initial conditions and models.

To meet the demands of these applications, we need to develop a fast solver that is capable of simultaneously simulating the dynamical system over a large  batch of initialization data.
The advances in processors, such as graphics processing units~(GPUs),\cite{brodtkorb2013graphics} tensor processing units,\cite{jouppi2017datacenter} and natural graphics processing units,\cite{tan2020fastva} provide  the possibility of accelerating the numerical computation via parallel computing of batch data. 
However, most hardware implements  restrictive SIMD-based (single instruction, multiple data) models.~\cite{liao2002high} The numerical method that needs individual processing of each trajectory is not appropriate for SIMD processors directly. 
For example, the adaptive time-step integrator 
(e.g., the Runge--Kutta-–Fehlberg method~\cite{fehlberg1969low})
determines a step size at each step based on an estimate of the local error, making the iterative computation in Eq.~(\ref{eq:iterative}) asynchronous for each trajectory in the batch and affecting the efficiency of parallel computing. On the other hand, we may control the step size to be the same at each step for all trajectories by adding one dimension to Eq.~\eqref{eq:iterative}, for example, $\mathbf{u}_{n+1}\in \mathbb{R}^{N\times d}$ with $N$ representing the batch size.
Controlling the step size requires considering a combined ODE system and estimating the error on all batch elements~\cite{chen2018neural}. 
The step size is limited by the largest local truncation error in a batch, making it difficult to use a large step size especially when the batch size is large.~\cite{chen2018neural,liang2022stiffnessaware} 
If the step size is always small in each step, it becomes slow for integration.
Therefore, SIMD processors prefer a fixed time-step integrator (i.e., $\Delta t:=\Delta t_1=\Delta t_2=\cdots$),
\begin{align}
     \mathbf{u}_{n+1} = \mathbf{u}_{n} + S(\mathbf{f}, \mathbf{u}_{n}, \Delta t), \quad \mathbf{u}_0 = \mathbf{c}_0, \quad n=0,1,\cdots .
     \label{eq:iterativefixed}
\end{align}
However, a fixed step size integrator encounters a trade-off\cite{li2021fourier,liang2022stiffnessaware} on step size between accuracy and computational efficiency: a large step size has a fast simulation but leads to a less accurate solution, while a small step size has a slow simulation but achieves a more accurate solution (see Table \ref{tab:table1} for the comparison of evaluation time and theoretical error between the traditional solvers with fine or coarse step size). This trade-off limits the feasibility of large-scale simulation if high accuracy is required.

To break through this speed-accuracy trade-off, in this paper we propose an open-source and data-driven corrector, called neural vector (NeurVec), which enables integration with coarse step size while maintaining the accuracy of fine step size in large-scale simulations. We emperically demonstrate that NeurVec is capable of overcoming the stability restriction of explicit integration methods for ODEs. The deployment of NeurVec comprises  offline training and inference (Fig.~\ref{fig:1}a). During offline training, NeurVec is trained with the accurate solution, while during inference NeurVec is employed to the solver to compensate for the error caused by the coarse step size. Our results on \rv{some complex dynamical system benchmarks} show that NeurVec is capable of learning the error term and accelerating the large-scale simulation of dynamical system significantly. Additionally, we have found that NeurVec not only overcomes the stability restriction of explicit integration methods for ODEs but also exhibits excellent generalization capabilities.

\rv{Some related works~\cite{pan2018long,liu2022predicting,chen2021generalized}
centered around purely data-driven methods for enhancing the accuracy of learning unknown dynamics from data. These approaches employ neural networks to characterize the system dynamics. For example, Ref.~\cite{pan2018long} improve long-term predictive modelling of nonlinear systems approximated by neural networks, by introducing supplementary regularization loss terms. In Ref~\cite{chen2021generalized}, neural networks are fused with existing coarse modeling techniques to improve the accuracy of a reduced/coarse model.
Different from the previous works that focus on enhancing the accuracy of learning unkown dynamics~\cite{dresdner2022learning}, our objective is to accelerate the numerical simulation of multiple trajectories, each initialized differently, for dynamics that are already known. In our approach, we use neural networks to compensate  the time integration errors that arise because of large time step sizes.} There are some previous works also consider accelerating the solution estimation via deep learning, but they emphasize  adopting pure data-driven approaches,~\cite{harlim2021machine,lu2021learning,li2021fourier} without using any explicit formula of the equation. Because of chaos\cite{choudhary2020physics,han2021adaptable,greydanus2019hamiltonian} (solution is sensitive to small perturbations) and stiffness\cite{liang2022stiffnessaware,huang1997adaptive} (solution is unstable unless a sufficiently small step size is used), the pure data-driven method still suffers from large errors in prediction, especially for long-term prediction\cite{wang2022stable}.

\section*{Framework of NeurVec}
In this study, we aim to discretize and solve the differential equation given by Eq.(\ref{eqn:ode}) using a larger step size, specifically $k$ times the step size ($k\Delta t$). To illustrate this approach, we consider the Euler method as an example, and use Taylor expansion to obtain:
\begin{equation}
\mathbf{u}(t+k\Delta t) = \underbrace{\mathbf{u}(t) + \mathbf{f}[\mathbf{u}(t)]\cdot k\Delta t}_{\text{For Euler method}} + \sum_{n=2}^\infty \underbrace{ \frac{1}{n!} \frac{\text{d}^n}{\text{d}t^n}\mathbf{u}(t)\cdot [k\Delta t]^n}_{\text{Error term err}_n }.
\label{eq:taylorneurvec}
\end{equation}
By neglecting all high-order error terms $\text{err}_n$ and discretizing $\mathbf{u}$, we can obtain the corresponding Euler method. However, when using a large step size of $k\Delta t$, discarding all high-order error terms will significantly reduce the accuracy of the Euler method, leading to poor predictions.  Notably, from Eq.(\ref{eqn:ode}), we have $\text{err}_n (k,\Delta t, \mathbf{u}(t)) \triangleq \frac{1}{n!} \frac{\text{d}^n}{\text{d}t^n}\mathbf{u}(t)\cdot [k\Delta t]^n = \frac{1}{n!} \frac{\text{d}^{n-1}}{\text{d}t^{n-1}}\mathbf{f}[\mathbf{u}(t)]\cdot [k\Delta t]^n$, which depends on $\mathbf{u}(t)$ and the constants $k$ and $\Delta t$. 
Therefore, we consider utilizing a corrector called NeurVec by a neural network with input $\mathbf{u}(t)$ to approximate $\sum_{n=2}^\infty \text{err}_n$. In other words, the corrector NeurVec, in the form of a vector function, is directly added to the estimated solution to compensate for the error caused by the use of the coarse step size (Fig.~\ref{fig:1}b). 
Following this idea, other forward numerical solvers besides the Euler method can also benefit from NeurVec to compensate for errors.
Specifically, NeurVec, a neural network parameterized by $\Theta$, maps from the state $\mathbb{R}^d$ to the error correction $\mathbb{R}^d$ (see the Methods section) and is added to the iterative formula of the solver with the step size $k\Delta t$ to get the solution $\{\hat{\mathbf{u}}_{kn}\}_{k=0}^\infty$, i.e.,
 \begin{align}
     \hat{\mathbf{u}}_{k(n+1)} = \hat{\mathbf{u}}_{kn} + S(\mathbf{f}, \hat{\mathbf{u}}_{kn}, k\Delta t) + \text{NeurVec}(\hat{\mathbf{u}}_{kn};\Theta), \quad \hat{\mathbf{u}}_0 = \mathbf{c}_0, \quad n=0,1,\cdots .
     \label{eq:iterative2}
 \end{align}
With NeurVec, we just need to estimate the solution on every $k$ steps instead of step by step as in Eq.~\eqref{eq:iterativefixed}. NeurVec is trained from the more accurate solutions with fine step size $\Delta t$ to characterize the error caused by the use of the coarse step size $k\Delta t$. The parameter $\Theta$ in NeurVec can be optimized by minimizing the mean squared difference between the predicted error and the error of the solver with the coarse step size: 
 \begin{align}
     \min_{\Theta}\frac{1}{G}\sum_{n=1}^G\big\|\text{NeurVec}(\mathbf{u}_{kn};\Theta)-\big(\mathbf{u}_{k(n+1)}-\mathbf{u}_{kn} - S(\mathbf{f}, \mathbf{u}_{kn}, k\Delta t)\big)\big\|_2^2,
     \label{eq:objective}
 \end{align}
where $G$ is the number of training samples $\{\mathbf{u}_s\}_{s=1}^{(G+1)k}$ from the fine step size. Table \ref{tab:table1} displays a comparison of evaluation time and theoretical error between the traditional solver and NeurVec. We  use $\epsilon$ to denote the runtime ratio of NeurVec to the scheme $S$. NeurVec inevitably increases the relative time complexity for each step by $\epsilon$ since an additional computation module is used \rv{(see supplementary material for details)}. When $k> (1+\epsilon)$, NeurVec with the coarse step size $k\Delta t$ is faster than the solver with the fine step size $\Delta t$, while achieving  comparable accuracy. Moreover, the runtime increment $\epsilon$ of NeurVec can be lessened. For example, the more complicated scheme $S$ increases the time complexity, and built-in parallel computing in Pytorch,\cite{paszke2019pytorch} TensorFlow,\cite{abadi2016tensorflow} or other deep learning frameworks enables smaller time complexity of NeurVec. As we will see in the Results  section, $\epsilon < 1$ uniformly, so NeurVec can accelerate the solver when $k\geq 2$. To characterize the solution error of NeurVec, we consider the Euler method, a simple ODE solver, as a proof of concept. The global truncation error of the Euler method linearly grows with the step size,\cite{butcher2016numerical} namely, $\mathcal{O}(\Delta t)$ when the step size is $\Delta t$ and $\mathcal{O}(k\Delta t)$ when the step size is $k\Delta t$. In our theory, we show that NeurVec of sufficient width can achieve an error of $\mathcal{O}(\Delta t)$ when the step size is $k\Delta t$, which breaks the accuracy-speed trade-off.

\begin{table}[htbp]
  \centering
  
    \begin{tabular}{llll}
    \toprule
    \multicolumn{1}{c}{Method} & \multicolumn{1}{c}{Step size} & \multicolumn{1}{c}{Evaluation time} & \multicolumn{1}{c}{Theory error (Euler scheme)} \\
    \midrule
    Fixed step size solver (fine step size) &  $\Delta t$     &  $\mathcal{O}(1/\Delta t)$      &  $\mathcal{O}(\Delta t)$\\
    Fixed step size solver (coarse step size) &  $k\Delta t$     &   $\mathcal{O}(1/(k\Delta t))$    &  $\mathcal{O}(k\Delta t)$\\
    NeurVec (coarse step size) & $k\Delta t$      & $\mathcal{O}((1+\epsilon)/(k\Delta t))$      &  $\mathcal{O}(\Delta t)$\\
    \bottomrule
    \end{tabular}%
  \caption{\label{tab:table1}Comparison of evaluation time and theoretical error (based on the Euler scheme) among the numerical solvers with fine $(\Delta t)$ or coarse step size $(k\Delta t)$ and NeurVec $(k\Delta t)$. Here $\epsilon$ denotes the ratio of the runtime of NeurVec to that of scheme $S$ for one step. 
  The fixed step size solvers suffer from the accuracy-speed trade-off on the step size. NeurVec learns from the solutions of fine step size. Then NeurVec is applied to the solver and integrates with the coarse step size ($k\Delta t$) but still has the theoretical accuracy of the fine step size, $\mathcal{O}(\Delta t)$. }
\end{table}%

\section*{Results}

We verify the capabilities of NeurVec in two aspects: (1) NeurVec is capable of stabilizing and accelerating the simulation on widely used numerical solvers with consistent performance improvement; and (2) NeurVec can be applied effectively to various \rv{complex dynamical system benchmarks.}

To illustrate the performance of NeurVec, we employ a simple network structure, a one-hidden-layer fully connected neural network~\cite{liang2022stiffnessaware}, to model NeurVec, where the number of the hidden neurons is 1,024 and a rational function~\cite{boulle2020rational} is used (see the Methods section for details). The training and inference of NeurVec are all performed on a single GeForce RTX 3080 GPU with a memory of 10 gigabytes. 
The simulations in the training and testing sets are uniformly sampled every time interval $\eta$, and hence we have discrete solutions on the time $0, \eta, 2\eta, \cdots$. To compare the speed, we average the clock time of simulations over 70 trials. 
The complete statistical tests can be found in the supplementary material.

In the preceding sections we used $\Delta^{\rm F}t$ and $\Delta^{\rm C}t$ to distinguish the fine step size from the coarse one used in traditional solver (e.g., $\Delta^{\rm F}t=\Delta t$ and $\Delta^{\rm C}t=k\Delta t$ in our previous notation). 
$\Delta^{\rm NV}t$ represents using NeurVec and integrating with step size $\Delta^{\rm NV}t$.

\begin{figure*}[t]
    \centering
    \includegraphics[width=\linewidth]{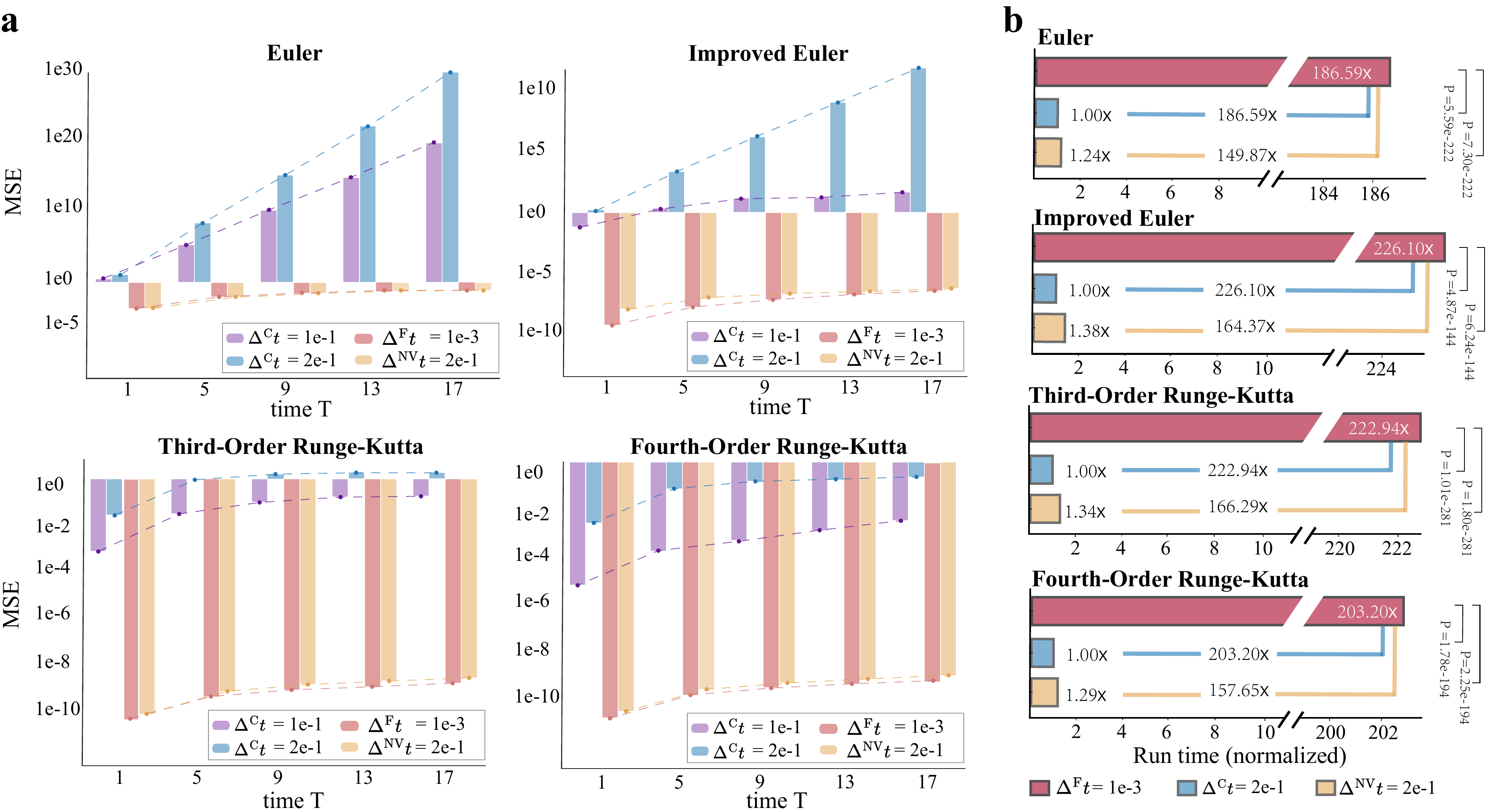}
    \caption{\textbf{Application of NeurVec on different numerical solvers.} \textbf{a}, The mean square error (MSE) between the reference solution and the numerical solutions with different configurations (step size $\Delta^{\rm C}t = 1e-1$, $\Delta^{\rm C}t=2e-1$, $\Delta^{\rm F}t=1e-3$, and NeurVec ($\Delta^{\rm NV}t = 2e-1$)) on the spring-chain system, averaged over 10.5k simulations. The reference solution is obtained by using the 4th-order Runge--Kutta with step size $1e-4$. NeurVec is trained on the simulations of $\Delta^{\rm F}t=1e-3$. The numerical solution of $\Delta^{\rm C}t=2e-1$ becomes unstable using Euler or the improved Euler formula, while NeurVec ($\Delta^{\rm NV}t=2e-1$) achieves a stable solution with  accuracy comparable to that of $\Delta^{\rm F}t=1e-3$. \textbf{b}, The (normalized) runtime of the numerical solver with $\Delta^{\rm F}t=1e-3$, $\Delta^{\rm C}t=2e-1$, and NeurVec ($\Delta^{\rm NV}t=2e-1$). The runtime of $\Delta^{\rm C}t=2e-1$ is benchmarked to one unit. NeurVec ($\Delta^{\rm NV}t=2e-1$) has  accuracy similar to  that  of $\Delta^{\rm F}t=1e-3$ and is over 150 times faster.}
    \label{fig:2}
\end{figure*}

\subsection*{NeurVec for \rv{ different numerical solvers.}}
We demonstrate the performance of NeurVec on widely used numerical solvers with consistent performance improvement. We perform NeurVec with four solvers:  Euler, improved Euler, third-order Runge--Kutta (RK3), and fourth-order Runge--Kutta (RK4) (see the Methods section) on a high-dimensional spring-chain system.\cite{Chen2020Symplectic} The spring-chain system we consider describes the motion of $d$ masses linked by $d+1$ springs, and the springs are placed horizontally with the two ends connected to a fixed wall. 
The ODE of the system is given by 
\begin{equation}
\frac{\text{d} q_i}{\text{d}t} = \frac{p_i}{m_i}, \quad \frac{\text{d} p_i}{\text{d}t} = k_i(q_{i-1}-q_i)+k_{i+1}(q_{i+1}-q_i), \quad i=1,2,\cdots, d, \quad q_0=q_{d+1}=0,
\label{eqn:springchain}
\end{equation}
where the variables $q_i$ and $p_i$ represent position and momentum of the $i$th mass, respectively, $i=1,\cdots, d$. Here $m_i$ and $k_i$ are the mass of the $i$th mass and force coefficient of the $i$th spring, respectively, and they are randomly and uniformly sampled (for exact value, see the supplementary material). 
We first introduce the training dataset to train NeurVec and the testing dataset for evaluation. The training and testing simulations are uniformly sampled every time interval $\eta = 2e-1$. The initial states are sampled randomly from uniform distribution $\pi:=\mathcal{U}([-2.5, 2.5]^d\times[-2.5, 2.5]^d)$. We set the dimension $d=20$ so the dimension of the state is $40$. Given a scheme $S$, the training dataset is generated by $S$ with $\Delta^{\rm F}t = 1e-3$. The reference simulations in the testing set are generated by RK4 with sufficiently small step size $1e-4$ (see the supplementary material).

Next, we demonstrate the performance of NeurVec in terms of accuracy and speed. NeurVec learns from the simulations of $\Delta^{\rm F}t=1e-3$ and is applied to the numerical solver with $\Delta^{\rm C}t=2e-1$. We characterize accuracy by the MSE between the reference and the simulated solution.  
In the short-term simulation on the time interval $[0,17]$, the numerical solutions of the coarse step size ($\Delta^{\rm C}t \geq 1e-1$) incur considerable simulation error and become unstable if Euler and the improved Euler are used (Fig.~\ref{fig:2}a). By contrast, NeurVec ($\Delta^{\rm NV}t=2e-1$) achieves a stable solution with  accuracy comparable to that of the fine step size $\Delta^{\rm F}t=1e-3$ (Fig.~\ref{fig:2}a), which means that NeurVec can overcome the stability restriction. 
These observations indicate that NeurVec learns the error distribution from the fine step size dataset and compensates for errors caused by the use of the coarse step size, demonstrating that  NeurVec is  compatible to these solvers. To better display the runtime, we benchmark the runtime of $\Delta^{\rm C}t=2e-1$ as one unit. The use of NeurVec increases
for a certain runtime ($\epsilon\leq 0.38$) for a single step (compared with $\Delta^{\rm C}t=2e-1$), but NuerVec has  accuracy comparable to that of $\Delta^{\rm F}t=1e-3$, which needs 200 steps for integrating over the time interval $2e-1$. The runtime of $\Delta^{\rm F}t=1e-3$ is much higher than that of NeurVec ($\Delta^{\rm C}t=2e-1$) ($P$ value $\ll$ 0.001 under two-sided t-tests), and NeurVec enables these numerical methods to have more than 150$\times$ speedup on the spring-chain systems (Fig.~\ref{fig:2}b).

\subsection*{NeurVec on \rv{complex} dynamical systems.}
We verify the effectiveness of NeurVec on challenging \rv{complex systems, including the classical chaotic system Hénon--Heiles system, elastic pendulum, and $K$-link pendulum. For more challenging systems, such as the Kuramoto-Sivashinsky equation (KSE), please refer to the supplementary materials.}. The chaotic system is sensitive to perturbation to the initial state, and small errors are increased exponentially by the dynamics. 
For all of these examples, we generate the testing set using RK4 with step size $1e-4$ while the training set is generated with $\Delta^{\rm F}t=1e-3$. The initial conditions are randomly and uniformly sampled on a range of values (see the supplementary material). NeurVec is applied to RK4.

\begin{figure*}[t]
    \centering
    \includegraphics[width=0.9\linewidth]{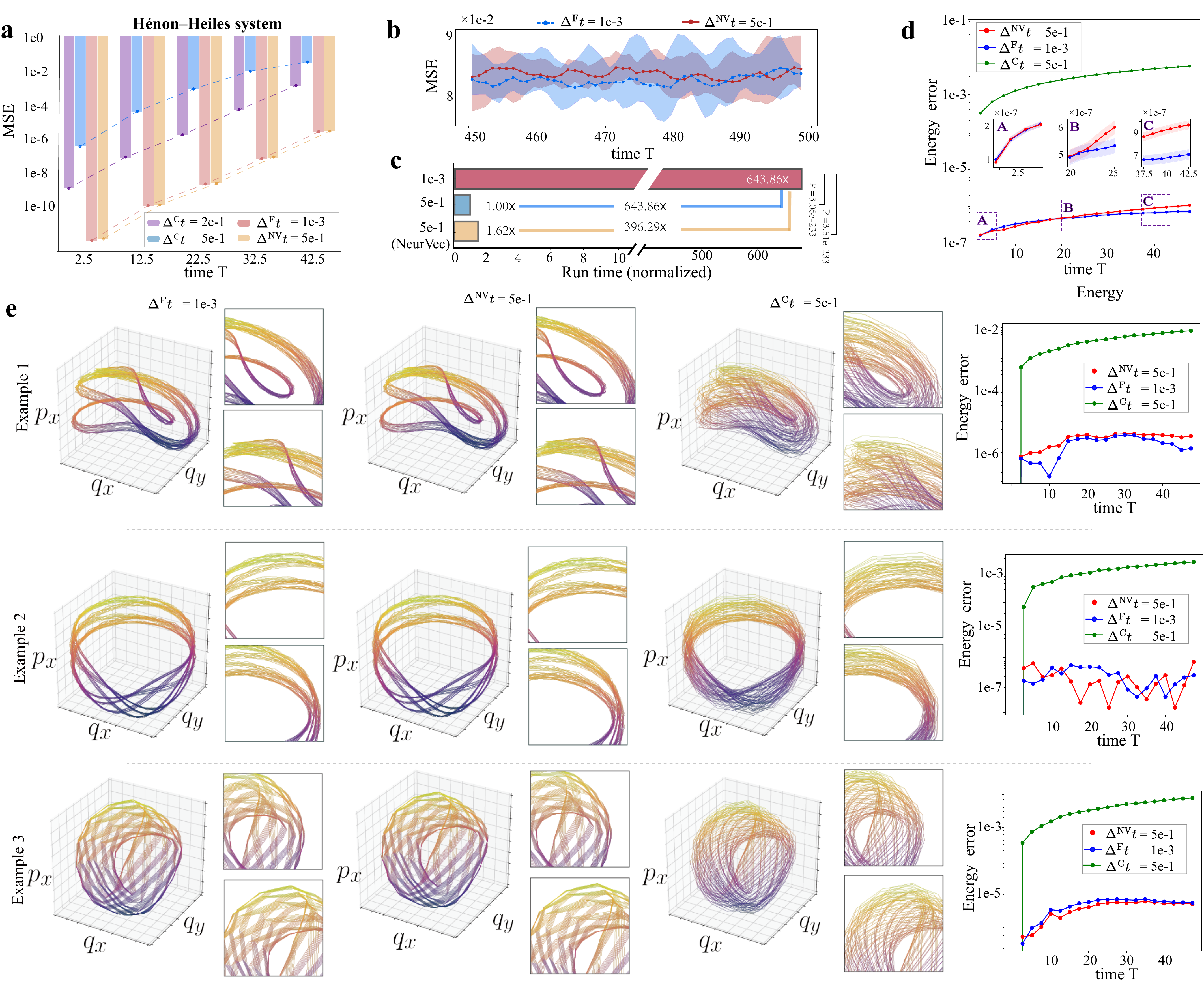}
    \caption{\textbf{Performance comparison on the Hénon--Heiles system.} \textbf{a}, MSE with varied time on the time interval $[0, 42.5]$ under different configurations (step size $\Delta^{\rm C}t = 5e-1$, $\Delta^{\rm C}t = 2e-1$, $\Delta^{\rm F}t=1e-3$, and NeurVec ($\Delta^{\rm NV}t = 5e-1$)). \textbf{b}, MSE with varied time on the longer time interval $[450, 500]$. The upper and lower bounds of the light color indicate the maximal and minimal error, respectively. \textbf{c}, The (normalized) runtime of the numerical solver with $\Delta^{\rm F}t=1e-3$, $\Delta^{\rm C}t=5e-1$, and NeurVec ($\Delta^{\rm NV}t=5e-1$). \textbf{d}, Energy error with varied time on $[0, 50]$. \textbf{e}, We provide three examples of trajectories projected on the coordinates $(q_x,q_y,p_x)$, and the corresponding energy error on $[0,50]$. More examples can be found in the supplementary material.}
    \label{fig:4}
\end{figure*}

\noindent \textbf{(1) Hénon--Heiles system}.
The Hénon--Heiles system is a Hamiltonian system\cite{feit1984wave} that describes the motion of a body around a center on the $x$-$y$ plane. Let $(q_x,q_y)$ and $(p_x,p_y)$ denote the positions and momenta of a particle, respectively. The ODE is given by
\begin{align}
    \frac{\text{d}}{\text{d}t}\left(q_x,q_y,p_x,p_y\right) = \left(p_x,p_y,-q_x-2\lambda q_xq_y, -q_y-\lambda(q_x^2-q_y^2)\right).
    \label{eq:hh}
\end{align}
The Hamiltonian (energy) function $\mathcal{H}$, defined by
\begin{equation}
    \mathcal{H}\left(q_x,q_y,p_x,p_y\right) = \frac{1}{2}(p_x^2+p_y^2)+\frac{1}{2}(q_x^2+q_y^2) + \lambda(q_x^2q_y-\frac{q_y^2}{3}),
    \label{eq:energy}
\end{equation}
must be conserved during the time evolution.
This property is used as an additional metric to evaluate the accuracy of our method. We characterize the energy error by the absolute difference between the energy of the simulated trajectory and the initial energy. The datasets are generated with initial energy between $[\frac{1}{12}, \frac{1}{6}]$ and $-1 < q_x < 1$, $-0.5 < q_y < 1$ such that the equipotential curves of the system form an inescapable interior region and exhibit chaotic behavior~\cite{NEURIPS2020_439fca36}. The simulations are uniformly sampled every time interval $\eta = 5e-1$. 

We find that NeurVec ($\Delta^{\rm NV}t=5e-1$) vastly improves the accuracy of the ODE solvers, achieves almost the same accuracy as RK4 with $\Delta^{\rm F}t=1e-3$ on the time interval $[0,42.5]$ (Fig.~\ref{fig:4}a), and works well for a much larger time interval $[450, 500]$ (Fig.~\ref{fig:4}c). Furthermore, NeurVec with $\Delta^{\rm NV}t=5e-1$ almost maintains the same system energy as does the reference method with $\Delta^{\rm F}t=1e-3$ (Fig.~\ref{fig:4}d). To illustrate the error correction capability of NeurVec, we visualize three trajectories of the first three components $(q_x,q_y,p_x)$ in Fig.~\ref{fig:4}e. For Examples 1--3 of Fig.~\ref{fig:4}e, NeurVec with $\Delta^{\rm NV}t=5e-1$ produces orbits similar to those of the reference method with $\Delta^{\rm F}t=1e-3$ while having an energy error of the same magnitude. Furthermore, the reference method with $\Delta^{\rm C}t=5e-1$ yields a larger energy error and pathwise difference. 
 Integrating with $\Delta^{\rm F}t=1e-3$ over the time interval $\eta = 5e-1$ takes $500$ steps, so it is not surprising that the runtime for $\Delta^{\rm F}t=1e-3$ is much larger than NeurVec with $\Delta^{\rm NV}t=5e-1$. Based on our test, NeurVec ($\Delta^{\rm NV}t=5e-1$) reaches more than 390$\times$ speedup over the reference method with $\Delta^{\rm F}t=1e-3$ (Fig.~\ref{fig:4}b). \\

\begin{figure*}[t]
    \centering
    \includegraphics[width=\linewidth]{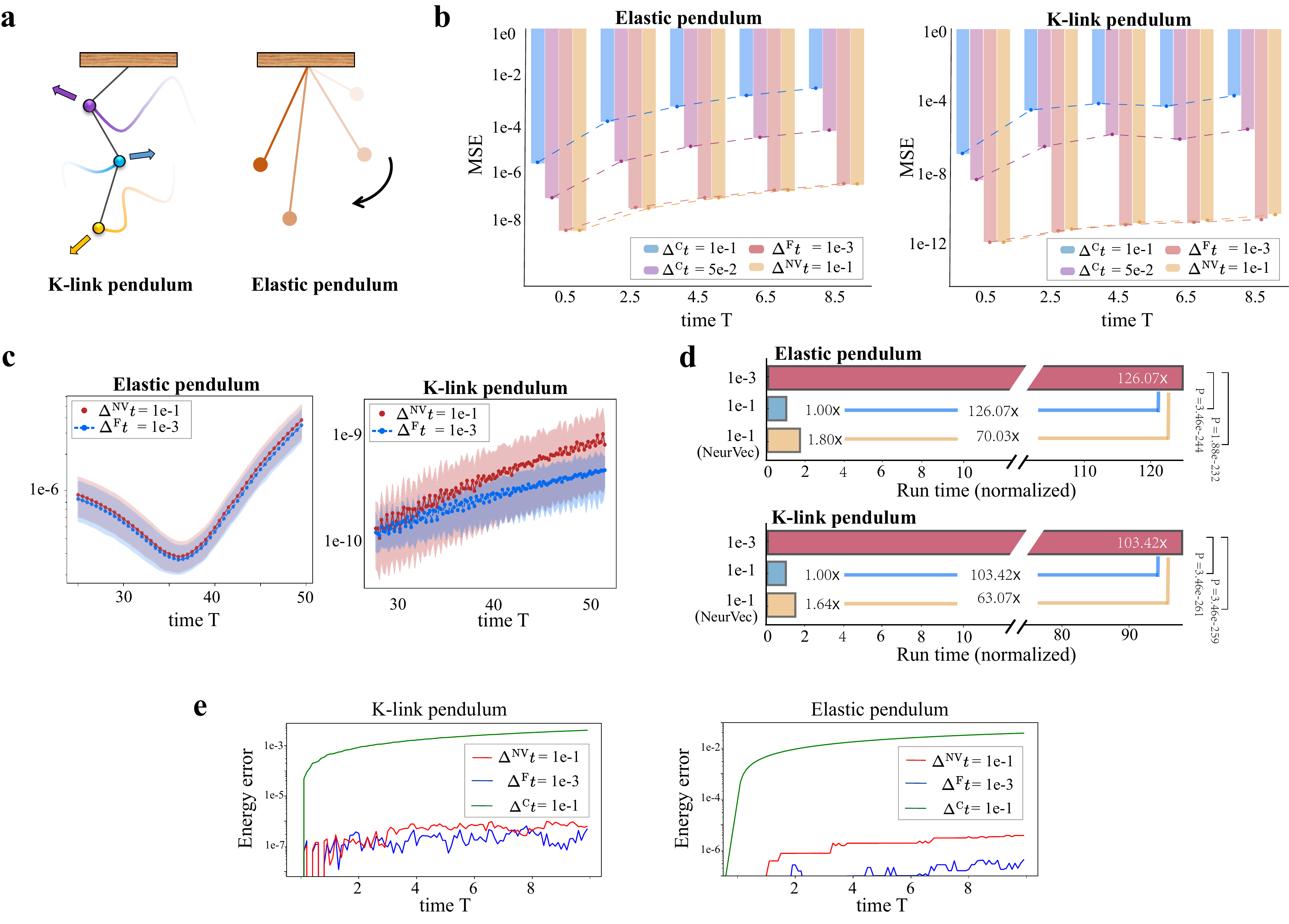}
     \caption{\textbf{Performance comparison on the pendulum systems.} \textbf{a}, The pendulum systems.  \textbf{b}, MSE with varied time on the time interval $[0, 8.5]$. \rv{\textbf{c}, MSE with varied time on the longer time interval $[25, 50]$. \textbf{d}, The runtime comparison among the numerical solvers with $\Delta^{\rm F}t=1e-3$, $\Delta^{\rm C}t=1e-1$, and NeurVec ($\Delta^{\rm NV}t=1e-1$). \textbf{e}, The energy error of two pendulum systems. } }
    \label{fig:5}
\end{figure*}

\noindent \textbf{(2) Elastic pendulum}. The elastic pendulum describes a point mass connected to a spring swinging freely (Fig.~\ref{fig:5}a), which may exhibit chaotic behavior under the force of gravity and spring~\cite{breitenberger1981elastic}. We denote $\theta$ as the angle between the spring and the vertical line and $r$ as the length of the spring. $\dot{\theta}$ and $\dot{r}$ correspond to the time derivative of $\theta$ and $r$, respectively. The motion of this system is governed by the ODE,
\begin{equation}
    \frac{\text{d}}{\text{d}t}\left(\theta,r,\dot{\theta},\dot{r} \right) = \left( \dot{\theta},\dot{r},\frac{1}{r}(-g \sin\theta-\dot{\theta}\dot{r}),r\dot{\theta}^2-\frac{k}{m}(r-l_0)+g\cos\theta \right),    \label{eq:elastic}
\end{equation}
where $k, m, l_0$, and $g$ are spring constant, mass,  original length, and gravity constant, respectively. The initial length of $r$ is $r(0) = l_0 = 10$, $\dot{r}$ and $\dot{\theta}$ are initialized by constant 0, and $\theta$ is randomly sampled from the uniform distribution $\mathcal{U}([0,\frac{\pi}{8}])$. The simulations in the training and testing sets are uniformly sampled every time interval $\eta=1e-1$. NeurVec is trained on the simulation generated by $\Delta^{\rm F}t=1e-3$. NeurVec ($\Delta^{\rm NV}t=1e-1$) has  accuracy of the same order as does $\Delta^{\rm F}t=1e-3$ on both short-term prediction (time interval $[0, 8.5]$) (Fig.~\ref{fig:5}b) and long-term prediction (time interval $[25, 50]$) (Fig.~\ref{fig:5}d). NeurVec ($\Delta^{\rm NV}t=1e-1$) is much faster than $\Delta^{\rm F}t=1e-3$ ($P$ value $\ll$ 0.001 under two-sided t-tests), reaching about 70$\times$ speedup (Fig.~\ref{fig:5}c).\\

\noindent \textbf{(3) $K$-link pendulum}.
 A $K$-link pendulum is a body suspended from a fixed point (Fig.~\ref{fig:5}a) with $K$ rods and $K$ bobs so that the body can swing back and forth under gravity.\cite{lopes2017dynamics} The system exhibits chaotic behavior. For simplification, the length of each rod and the mass of each bob are set to 1, and the gravity constant $g$ is set to 9.8. Let variables $\bm{\theta}:= (\theta_1,\theta_2,\cdots,\theta_K)$, where $\theta_i$ is the angle between the $i$th rod and the vertical axis. The system is governed by the ODE
\begin{equation}
    \frac{\text{d}}{\text{d}t}(\bm{\theta},\dot{\bm{\theta}}) =  (\dot{\bm{\theta}},\mathbf{A^{-1}}\mathbf{b}).
    \label{eq:klink}
\end{equation}
Here $\mathbf{b}=(b_1,b_2,\cdots,b_K)$ and $b_i=-\sum_{j=1}^{K}\left[c(i, j) \dot{\theta}_{j}^{2} \sin \left(\theta_{i}-\theta_{j}\right)\right]-(K-i+1) g\sin \theta_{i}$.
 $\mathbf{A}$ is a $K \times K$ matrix with $\mathbf{A}_{i, j}=c(i, j) \cos \left(\theta_{i}-\theta_{j}\right)$, where $c(i,j) = K - \max(i,j)+1$, for $i,j=1,\cdots,K$. 
We use the example of a 2-link pendulum ($K=2$) to verify the advantage of NeurVec in terms of efficiency. 
The trajectory $(\bm{\theta},\dot{\bm{\theta}})$ is a 4-dimensional vector. The initial conditions  $\bm{\theta}(0)$ are randomly sampled on $\mathcal{U}[0,\pi/8]$, and $\dot{\bm{\theta}}$ are set to zero (see the supplementary material). NeurVec ($\Delta^{\rm NV}t = 1e-1$) significantly improves the accuracy of $\Delta^{\rm C}t = 1e-1$, and achieves  accuracy similar to that of RK4 ($\Delta^{\rm F}t = 1e-3$) for both short-term and long-term prediction (Fig.~\ref{fig:5}b and Fig.~\ref{fig:5}c). Moreover, NeurVec has over 63$\times$ speedup (Fig.~\ref{fig:5}d). \rv{Furthermore, Fig.~\ref{fig:5}e illustrates the energy error over time. We can see that NeurVec effectively conserves energy in simulations of a $K$-link pendulum and an elastic pendulum.}

\begin{figure*}[t]
    \centering
    \includegraphics[width=\linewidth]{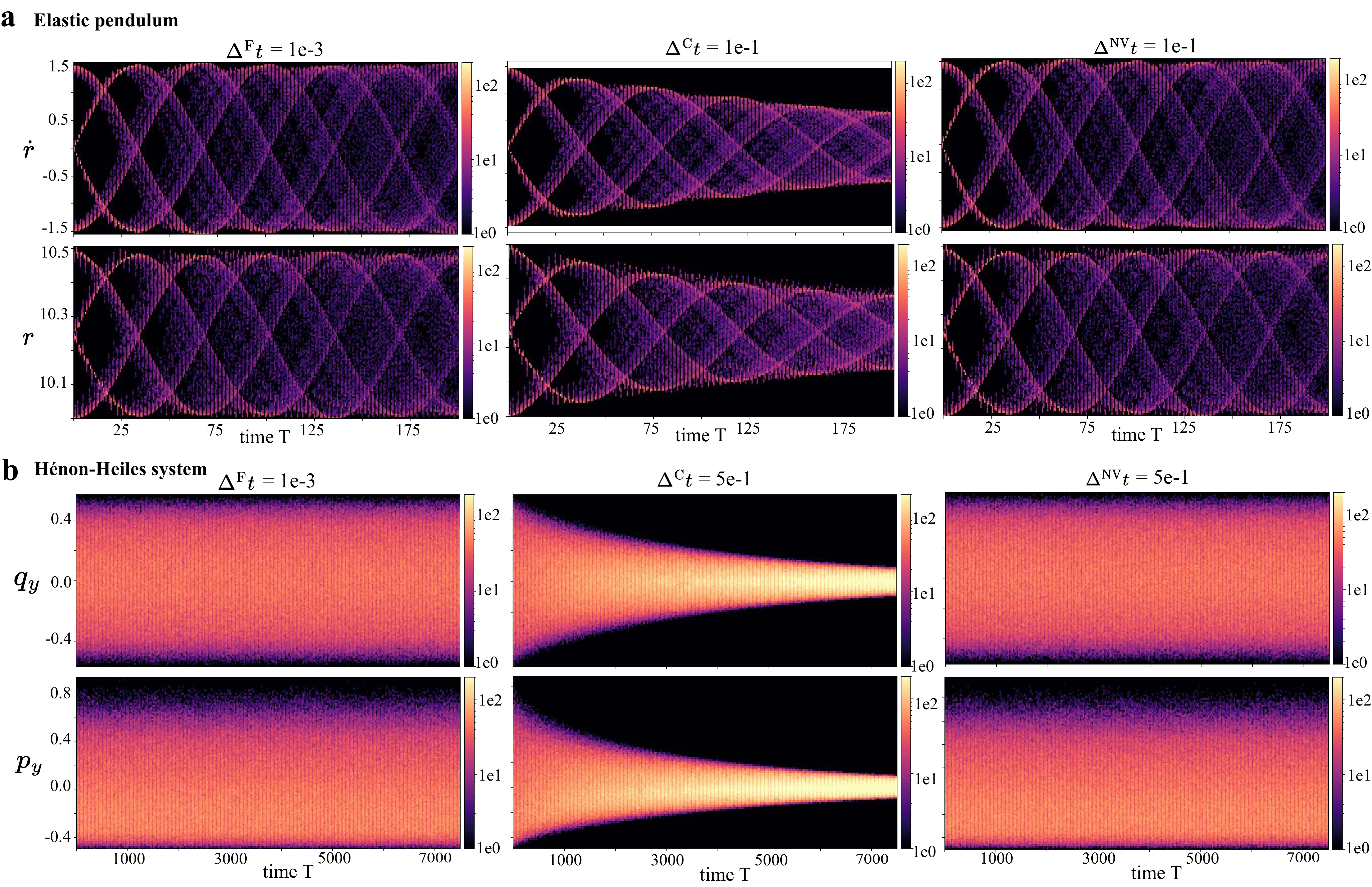}
    \caption{\textbf{Time series histogram.} We visualize the time series histogram of a test set for variables (\textbf{a}) $r$ and $\dot{r}$ in the elastic pendulum~\eqref{eq:elastic} and (\textbf{b}) $q_y$ and $p_y$ in the Hénon--Heiles system~\eqref{eq:hh}. The color represents the number count (the lighter color and the larger frequency). The solutions generated by the solver with coarse step size exhibits a trend of convergence to a specific value, while solutions of the solver with fine step size are distributed within a range, and NeurVec with coarse step size produces a histogram  visually identical with that of the solver with fine step size. This result shows that NeurVec has a more accurate solution than does the solver with fine step
    size. }
    \label{fig:6}
\end{figure*}
\section*{Analysis}

In this section we provide further analysis of the performance of NeurVec from three aspects: (1) NeurVec maintains the statistic of the solutions from large-scale simulation, which is crucial for constructing reliable models; (2) NeurVec learns the leading error term of the numerical solver, which enables a more accurate estimation for each step; and (3) we compare the evaluation time and solution error among solver with fine or coarse step size and NeurVec.\\

\noindent\textbf{Maintaining solution statistics in large-scale simulations}.

\noindent We validate the performance of NeurVec on producing  consistent statistical observations for ensemble forecasting. 
The ability to enable a large step size for a set of sampled initial conditions is critical for real applications such as weather forecasting. We visualize the time series histogram of the testing set for variables (\textbf{a}) $r$ and $\dot{r}$ in the elastic pendulum~\eqref{eq:elastic} and (\textbf{b}) $q_y$ and $p_y$ in the Hénon--Heiles system~\eqref{eq:hh} in Fig.~\ref{fig:6}. The time series histogram is generated by dividing axes into $800\times 100$ bins and counting the curves that cross the bins.

For the elastic pendulum, we find that starting from a time $T\geq 25$, the statistical difference of  $\dot{r}$ and $r$ between $\Delta^{\rm F}t=1e-3$ and $\Delta^{\rm C}t=1e-1$ becomes larger. 
When the step size is $\Delta^{\rm F}t=1e-3$, $\dot{r}$ exhibits periodic behavior, which is in accordance with the periodic variation of the spring during its extend-retract. However, let $\Delta^{\rm C}t=1e-1$, $\dot{r}$ and $r$ show a trend of approaching specific values, and the change range gradually narrows. On the other hand, the simulations with NeurVec ($\Delta^{\rm NV}t=1e-1$) have a pattern smilar to that of  $\Delta^{\rm F}t=1e-3$. We have a similar observation for $q_y$ and $p_y$ in Hénon--Heiles system~(Fig.~\ref{fig:6}b). Therefore, we conclude that NeurVec produces more accurate solutions compared with the reference method with large step size, enabling  better and more consistent statistical observation.\\

\noindent\textbf{Learnability and generalizability}. 

\noindent 
In Eq.(\ref{eq:taylorneurvec}), we start from the mathematical expression of the error term $\text{err}_n(k,\Delta t, \mathbf{u}(t))$ in the form of $\frac{1}{n!} \frac{\text{d}^n}{\text{d}t^n}\mathbf{u}(t)\cdot [k\Delta t]^n$, and consider using a neural network to approximate the sum of all error terms. However, neural networks are generally regarded as black box functions with a lack of interpretability. Therefore, in order to explain the good performance of NeurVec and confirm the learnability of the error terms, in this section, we explore and visualize what the neural network in NeurVec learns.
We consider solving a 1-link pendulum with the Euler method. Our consideration for testing NeurVec on this system is based on the following motivations. First, the dimension of the state is 2, which facilitates error visualization on phase space. Second, we derive the error term of the Euler method explicitly through the Taylor formula:
 \begin{equation}
     \mathbf{u}(t+\Delta t) - \left(\mathbf{u}(t) + 
     \mathbf{f}(\mathbf{u})\Delta t \right)=  \frac{1}{2}(\nabla \mathbf{f})  \mathbf{f}(\mathbf{u})\Delta t^2 +\mathcal{O}(\Delta t^3),
     \label{eq:expan}
 \end{equation}
where $\nabla \mathbf{f}$ is the Jacobian matrix of $\mathbf{f}$. The second-order term $\frac{1}{2}(\nabla\mathbf{f}) \mathbf{f}(\mathbf{u})\Delta t^2$ is the leading error term of the Euler method, which is supposed to be captured by NeurVec from data of fine step size. 
\begin{figure*}[t]
    \centering
    \includegraphics[width=\linewidth]{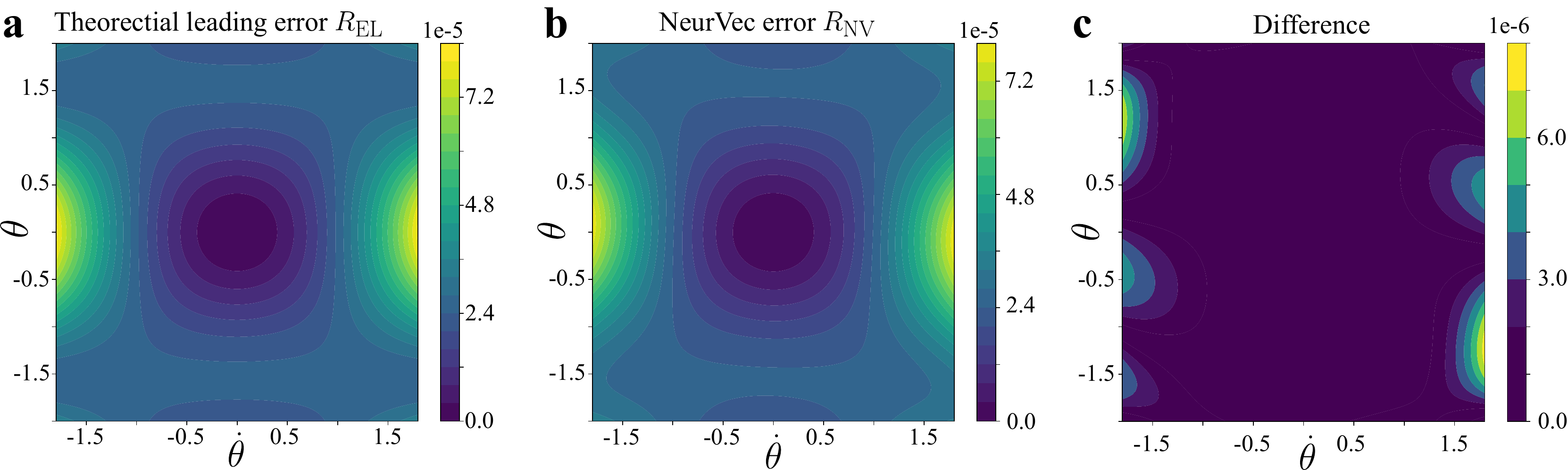}
    \caption{\textbf{Numerical error visualization on the phase space of 1-link pendulum.} \textbf{a}, The square sum of leading (second-order) error term of the Euler method, denoted by $R_{\rm EL}$. The error is calculated by using the true dynamics ${f}$. \textbf{b}, The square sum norm of error compensation learned by NeurVec, denoted by $R_{\rm NV}$. \textbf{c}, The difference between the leading error term and NeurVec. }
    \label{fig:theory}
\end{figure*}
Denote $R_{\text{NV}}(\mathbf{u}):=\left\|\text{NeurVec}(\mathbf{u})\right\|_2^2$ as the norm of error learned by NeurVec and  $R_{\text{EL}}(\mathbf{u}):=\left\|\frac{1}{2}(\nabla \mathbf{f}) \left( \mathbf{f}(\mathbf{u}\right)\Delta t^2\right\|_2^2$ as the norm of the leading error term of the Euler method. 
To train NeurVec for the Euler method, we generate the dataset by randomly sampling the initial conditions of $\theta$ and $\dot{\theta}$ from uniform distributions $\mathcal{U}([0,\pi/2])$ and $\mathcal{U}([0,0.5])$, respectively, and then use the Euler method to simulate the data with $dt = 1e-3$. We train  NeurVec with coarse step size $dt=1e-1$. 

We found that the learned error $R_{\text{NV}}$ (Fig.~\ref{fig:theory}a) is visually consistent with the leading error $R_{\text{EL}}$ of the Euler method (Fig.~\ref{fig:theory}b). The squared difference $R_{\text{Diff}}=\left\|\frac{1}{2}(\nabla \mathbf{f}) \left(\mathbf{f}(\mathbf{u}\right)\Delta t^2-\text{NeurVec}(\mathbf{u})\right\|_2^2$ is up to order $\mathcal{O}(10^{-6})$, and only a small part of the difference near the boundary is relatively large (Fig.~\ref{fig:theory}c). \rv{The relatively significant differences near the boundary may be attributed to the limited availability of data that encompasses those boundary regions. (see the supplementary material for details)}. Through the training data of high accuracy, NeurVec captured the leading error term of the numerical solver.  NeurVec may even capture the higher-order error terms, which enables the  use of a coarse step size. Furthermore, we can analyze the generalization ability of NeurVec from Fig.\ref{fig:theory}. On one hand, the results displayed in Fig.~\ref{fig:theory}c demonstrate that NeurVec is capable of generalizing well on a continuous phase space, even when trained using limited and discrete data. This suggests that NeurVec exhibits good interpolation generalization abilities.
 On the other hand, it is noteworthy that the initial conditions of our training data were sampled from $\mathcal{U}([0,\pi/2])$ and $\mathcal{U}([0,0.5])$, and the testing range shown in Fig.~\ref{fig:theory} shows the neural network's effective approximation of the error term and implicit prior information, indicating a certain degree of extrapolation capability. Such findings suggest that NeurVec demonstrates a promising potential for generalizing beyond the training data.\\

\noindent\textbf{Theoretical analysis}. We analyze the runtime and the global error in the solution approximated with NeurVec. Let $0=t_0<t_1<\cdots<t_{pk}=T$ be uniform points on $[0, T]$ and $\Delta t=\frac{T}{pk}$. 

We compare the runtime of fine and coarse step size. If the step size $\Delta t$ is used, then the number of steps for integration is $\frac{T}{\Delta t}$. If the step size is $k\Delta t$, then the number of steps needed is $\frac{T}{k\Delta t}$. $\epsilon$ is the ratio of the runtime of NeurVec to that of scheme $S$ for one step.
Hence, when NeurVec is used to integrate with step size $k\Delta t$, we need $\epsilon\times 100\%$ extra time for each step; and the time becomes $\mathcal{O}(\frac{T(1+\epsilon)}{k\Delta t})$.

Next we study the error of solvers with fine or coarse step size. For simplification, we focus on the Euler solver and characterize the global discretization error (difference between the true solution and the estimated solution) at the time $T$. When the Euler scheme is used, \begin{align}
    \mathbf{u}_{n+1} = \mathbf{u}_{n} + \Delta t \mathbf{f}(\mathbf{u}_{n}), \quad \mathbf{u}_0 = \mathbf{c}_0, \quad n=0,1,\cdots, kp-1. 
    \label{eqn:eulertheory}
\end{align}

\begin{Proposition}
\label{thm:euler}
We assume that (1) $\mathbf{f}$ is Lipschitz continuous with Lipschitz constant $L$ and (2) the second derivative of the true solution $\mathbf{u}$ is uniformly bounded by $M>0$, namely., $\|\mathbf{u}''\|_\infty\leq M$ on $[0,T]$. Then, using~\eqref{eqn:eulertheory}, we have 
\begin{align*}
    |\mathbf{u}_{kp}-\mathbf{u}(T)|\leq \frac{M\exp(2TL)}{2L}\Delta t .
\end{align*}
\end{Proposition}
For the proof, see 
the work of Atkinson et al.~\cite{atkinson2011numerical}. Proposition~\ref{thm:euler} shows that the Euler method converges linearly. The error is $\mathcal{O}(\Delta t)$ when the step size $\Delta t$ is used. Similarly, the error becomes $\mathcal{O}(k\Delta t)$ when the step size is $k\Delta t$. We next derive the error for the Euler method with step size $k\Delta t$ using NeurVec. The iterative formula is given by 
\begin{align}
     \hat{\mathbf{u}}_{k(n+1)} = \hat{\mathbf{u}}_{kn} + f(\hat{\mathbf{u}}_{kn}) (k\Delta t) + \text{NeurVec}(\hat{\mathbf{u}}_{kn};{\Theta}), \quad \hat{\mathbf{u}}_0 = c_0, \quad n=0,1,\cdots, p-1. 
     \label{eq:iterative_neurvec}
\end{align}
We can use the following loss function to identify the learnable parameter $\Theta$ in NeurVec. $V_n$ denotes the residual error for each term. 
\begin{align}
 \text{LS} = \frac{1}{p}\sum_{n=0}^{p-1}\big\|\frac{\mathbf{u}_{k(n+1)}-\mathbf{u}_{kn}}{k\Delta t} - \mathbf{f}(\mathbf{u}_{kn}) - \frac{\text{NeurVec}(\mathbf{u}_{kn};{\Theta})}{k\Delta t}\|_2^2:=\frac{1}{p}\sum_{n=0}^{p-1}\|V_n\|_2^2
 \label{eq:objective_euler}
\end{align}
In the next theorem we characterize the error of NeurVec ($k\Delta t$) by the quality of the training data and the neural network training error. In addition to the assumption in Proposition~\ref{thm:euler}, we assume NeurVec is Lipschitz continuous and the Lipschitz constant is of order $k\Delta$. This assumption is reasonable based on the following motivation. According to Taylor expansion $v(t+\Delta t)=v(t)+v'(t)\Delta t+o(\Delta t)$, from our objective we expect that $\text{NeurVec}\sim o(k\Delta t)$.
\begin{theorem} Assumptions (1) and (2) in Proposition~\ref{thm:euler} hold. In addition, we assume that $\text{NeurVec}$ is Lipschitz continuous with Lipschitz constant $k\Delta t L_{NV}$, which is independent of $\bm{\theta}$. Then the error is 
\begin{align}
     |\hat{\mathbf{u}}_{kp}-\mathbf{u}(T)|\leq \frac{M\exp(2TL)}{2L}\Delta t +  \frac{\sqrt{T}\exp(T(L+L_{NV}))}{\sqrt{L+L_{NV}}}(\text{\rm LS})^{\frac{1}{2}.}
     \label{ineq:inequality}
\end{align}

\end{theorem}

The first term in the right-hand side of \eqref{ineq:inequality} comes from the error of the training data, Euler simulation with step size $\Delta t$, while the second term is the training error of NeurVec. A series of works~\cite{du2018gradient,jacot2018neural,chizat2019lazy} utilize a neural tangent kernel to prove the global convergence of a neural-network-based least squares method. Under the assumptions of training data distribution, when the width of one hidden layer network is sufficiently large, gradient descent converges to a globally optimal solution for the quadratic loss function. We might assume that the training error $\text{LS}\to 0$ as the increasing update iteration. Then in \eqref{ineq:inequality}, $|\hat{\mathbf{u}}_{kp}-\mathbf{u}(T)|\sim \mathcal{O}(\Delta t)$.

\section*{Discussion}

To address the speed-accuracy trade-off in large-scale simulations of dynamical systems, we proposed NeurVec, a deep learning based corrector in the numerical solver, which can compensate for the error caused by the use of coarse step size for numerical solvers. Through extensive experiments and preliminary theoretical evidence, we show that NeurVec is general and can be applied to widely used explicit integration methods and learn the error distribution through simulations with fine step size. However, there are still some limitations while using NeurVec. 

Notice that, according to the analysis in Table \ref{tab:table1}, NeurVec can achieve fast simulation, i.e., a speedup of $\mathcal{O}(k/(1+\epsilon))$ times, under the condition that, after the neural network is trained, it can maintain the same accuracy as the training data at a sufficiently large step size. However, there exist some complex differential equations, such as ultra-high dimensional dynamical systems, may not satisfy this condition and prevent the term $\frac{\sqrt{T}\exp(T(L+L_{NV}))}{\sqrt{L+L_{NV}}}(\text{\rm LS})^{\frac{1}{2}}$ in Eq.(\ref{ineq:inequality}) from converging to 0, which would make NeurVec not necessarily achieve the target accuracy $\mathcal{O}(\Delta t)$. Specifically, we consider a general high-dimensional dynamical system with dimension $d$, and a shallow network structure with a width of $N_1$ as shown in Fig.\ref{fig:nnarc} as an example.
In addition, Lu et al \cite{lu2017expressive} reveal that for an infinitely deep network with input dimension $d$ and some mild assumptions, it can achieve an arbitrary accuracy while the width of network is at least $\mathcal{O}(d)$. Therefore, the dimension of differential equations that NeurVec can currently handle will not exceed $\mathcal{O}(N_1)$, and in practice, since the neural network depth is limited, the available dimension may be much lower than $\mathcal{O}(N_1)$. To improve the accuracy, the easiest way is to widen the neural network as much as possible or consider more advanced learning architectures and training algorithms, for example, an attention mechanism,\cite{hu2018squeeze,huang2020dianet,liang2020instance} neural network structure search,\cite{he2021blending,liu2018darts,huang2021rethinking}, and large-scale pretrained models\cite{kenton2019bert}. However, these improvements will increase $\epsilon$, and the speedup ratio $\mathcal{O}(k/(1+\epsilon))$ will become relatively small at this time, even accelerated simulations cannot be achieved while $\epsilon$ is large enough.

In Table \ref{tab:highdim}, we consider the Sping-chain system to explore the impact of ultra-high dimensional dynamical systems on NeurVec. In this paper, we set $N_1 = $ 1024, the experimental results show that as the dimension increases, the accuracy of NeurVec becomes harder to maintain, especially in the case that the dimension close to $\kappa$, which is consistent with our analysis.

\begin{table}[htbp]
  \centering
\resizebox{0.9\hsize}{!}{
    \begin{tabular}{lc|cc|cc|cc}
    \hline
    \multicolumn{1}{c}{\textbf{Method}} & \multicolumn{1}{c|}{\textbf{Step Size}} &\multicolumn{1}{c}{\textbf{Dim = 300}} & \multicolumn{1}{c|}{\textbf{P<0.05?}} & \multicolumn{1}{c}{\textbf{Dim = 500}} & \multicolumn{1}{c|}{\textbf{P<0.05?}} & \multicolumn{1}{c}{\textbf{Dim = 1000}} & \multicolumn{1}{c}{\textbf{P<0.05?}} \\
    \hline
    Euler & $1e-3$   &  $2.58e-2$  &  -     &   $7.32e-2$    & -       &  $5.48e-1$     & -  \\
    Euler+NeurVec & $2e-1$  & $2.58e-2$   &  $\times$      &  $7.33e-2$      & $\times$      & $5.54e-1$      & $\times$ \\
    RK4   & $1e-3$  &  $4.86e-7$   &  -     &   $4.64e-6$    & -      & $6.42e-5$      & - \\
    RK4+NeurVec & $2e-1$  & $4.84e-7$   &  $\times$     &  $1.05e-5$     &  \checkmark     & $2.61e-3$      & \checkmark \\
    \hline
    \end{tabular}%
    }
      \caption{The impact on accuracy (MSE at evaluation time $T=20$) under the ultra-high dimension systems. All experimental settings are following the experiments in Fig.\ref{fig:2}. The "Dim" is the dimension of the dynamical system (Spring-chain).  We also consider the student t-test with a significant level of 0.05 for NeurVec. If the p-value P < 0.05, there exists a significant difference between NeurVec and the corresponding traditional numerical solvers and it also means NeurVec can not maintain consistent accuracy.}
  \label{tab:highdim}%
\end{table}%

In addition, we may extend NeurVec in several ways. 
\begin{itemize}
    \item \textbf{Generalizing NeurVec}. We implemented NeurVec using the simulation data with fixed system parameters and fixed step size. Recently, the neural network, as a universal approximator, shows promising results on learning the nonlinear continuous operator. Motivated by operator learning,\cite{lu2021learning,li2021fourier} we may add additional dimensions to the input of NeurVec, such as the step size for integration and the physical parameters in the system. Then NeurVec can be trained with more diverse simulation data, such that NeurVec can be used with different $dt$ for systems with varied physical parameters (such as $\bm{k}$ and $\bm{m}$ in the spring-chain system~\eqref{eqn:springchain}).
    
    \item \textbf{Continual model update}. 
    Neural networks may sometimes have inaccurate predictions when encountering abnormal situations. Therefore, we may need to maintain and update NeurVec regularly to learn from new data. The simplest strategy is to retrain the network from scratch, but it needs considerable computing resources to train and memory resources to store the data. To address such a problem, we can fine-tune NeurVec via incremental learning\cite{wu2019large} for a small amount of newly collected training data to achieve low-cost model updates.
    \item \textbf{Numerical solvers and cutting-edge problems}. In this paper, we consider four forward numerical solvers and four kinds of ODE problems.
    NeurVec can be extended to other types of numerical solvers, such as backward methods and implicit methods. These methods are mainly aimed at improving the simulation accuracy, but the simulation cost for one step may be large.
    \rv{For more intricate cutting-edge problems, NeurVec may require further tuning of training parameters, neural network architecture, and data processing, as explored in Refs. \cite{liu2022predicting,dresdner2022learning,huang2023robust}. Nevertheless, the NeurVec framework we introduce is general and easily extendable. We believe NeurVec has the potential to be adopted for more complex dynamical systems.}

\end{itemize}

\section*{Methods}

 \noindent\textbf{Datasets}. We summarize the simulated training and testing datasets used in the main text in Table~\ref{tab:data1}. For each dataset we integrate with the step size $\delta$ using the numerical solver over $N$ random initializations. We obtain the discrete solutions every $\delta$ up to the model time $T$. Next, we sample the solution every time interval $\eta$ ($\eta$ is a multiple of $\delta$). 

\begin{table}[htbp]
  \centering
    \begin{tabular}{llccccc}
    \toprule
    \multicolumn{1}{c}{Problem} & \multicolumn{1}{c}{Type} & Dim & Num & Step size $\delta$ & Method & Duration $T$\\
    \midrule
    Spring-chain (Euler) & Train & 40    & 60k   & 1e-3  & Euler & 20 \\
    Spring-chain (Improved Euler) & Train & 40    & 60k   & 1e-3  & Improved Euler & 20 \\
    Spring-chain (RK3) & Train & 40    & 60k   & 1e-3  & RK3   & 20 \\
    Spring-chain (RK4) & Train & 40    & 60k   & 1e-3  & RK4   & 20 \\
    Spring-chain & Test  & 40    & 10.5k & 1e-4  & RK4   & 20 \\
    \midrule
    1-link pendulum & Train & 2     &  1k     & 1e-3  & RK4   &  10\\
    2-link pendulum & Train & 4     & 300k  & 1e-3  & RK4   & 10 \\
    2-link pendulum & Test  & 4     & 7k    & 1e-4  & RK4   & 10 \\
    \midrule
    Hénon--Heiles & Train & 4     & 100k  & 1e-3  & RK4   & 50 \\
    Hénon--Heiles & Test  & 4     & 70k   & 1e-4  & RK4   & 50 \\
    \midrule
    Elastic pendulum & Train & 4     & 300k  & 1e-3  & RK4   & 50 \\
    Elastic pendulum & Test  & 4     & 14k    & 1e-4  & RK4   & 50 \\
    \bottomrule
    \end{tabular}%
  \caption{\label{tab:data1}Summary of the simulated datasets used in the main text. }
\end{table}%

 \noindent\textbf{Numerical solvers}. We introduce four numerical solvers used in our paper:  the Euler method, improved Euler method, and 3rd- and 4th-order Runge--Kutta methods. These solvers have different $S(f, u_{n}, \Delta t_n)$ in the iterative formula~\eqref{eq:iterative}.
\noindent (1) The Euler method can be written as
\begin{equation}
S(\mathbf{f}, \mathbf{u}_{n}, \Delta t_n) = \Delta t_n \mathbf{f}(\mathbf{u}_{n}).
\end{equation}
It has an explicit geometric interpretation---it uses a series of line segments to approximate the solution of the equation. 
It is first-order accurate since its local truncation error is $\mathcal{O}(\Delta t^2)$ and the global error is $\mathcal{O}(\Delta t)$.

\noindent (2) The improved Euler method \cite{suli2003introduction,trench2013elementary} can be written as
\begin{equation}
S(\mathbf{f}, \mathbf{u}_{n}, \Delta t_n) =\frac{\Delta t_n }{2}[ \mathbf{f}(\mathbf{u}_{n}) + \mathbf{f}(\mathbf{u}_{n} + \Delta t_n \mathbf{f}(\mathbf{u}_{n}))].
\label{eq:improveeuler}
\end{equation}
The improved Euler method is a numerical method that uses an implicit trapezoidal formulation to improve the accuracy of the Euler method. Specifically,  it first takes a one-step Euler method to obtain $\tilde{u}_{n+1} = u_{n} + \Delta t_n f(u_{n})$ and then uses the implicit trapezoidal formula to obtain $ u_{n+1} = u_{n} + \frac{\Delta t_n }{2}[f(u_{n}) + f(\tilde{u}_{n+1})]$. Even though the improved Euler method requires more computation compared with the Euler method, it has a higher accuracy with a local error of $\mathcal{O}(\Delta t^3)$. 
\noindent (3)  The $m$th-order Runge--Kutta method can be written as
\begin{equation}
S(\mathbf{f}, \mathbf{u}_{n}, \Delta t_n) = \Delta t_n \sum_{i=1}^m \lambda_iK_i.
\label{eq:rk}
\end{equation}
For the 3rd-order RK method, the number of stages $m=3$, the coefficients $\lambda_1 = \lambda_3 = \frac{1}{6}$ and $\lambda_2 = \frac{2}{3}$, and the update rule $K_1 = \mathbf{f}(\mathbf{u}_{n}), K_2 = \mathbf{f}(\mathbf{u}_{n}+\frac{\Delta t_n}{2}K_1)$ and $K_3 = \mathbf{f}(\mathbf{u}_{n} - \Delta t_nK_1+2\Delta t_nK_2)$. For the 4th-order RK method, $m=4$, $\lambda_1 = \lambda_4 = \frac{1}{6}$ and $\lambda_2 = \lambda_3 = \frac{1}{3}$. The update rule $K_1 = \mathbf{f}(\mathbf{u}_{n}), K_2 = \mathbf{f}(\mathbf{u}_{n}+\frac{\Delta t_n}{2}K_1), K_3 = \mathbf{f}(\mathbf{u}_{n}+\frac{\Delta t_n}{2}K_2)$ and $K_4 = \mathbf{f}(\mathbf{u}_{n}+\Delta t_nK_3)$.
Runge--Kutta methods, especially the 4th-order Runge--Kutta method, are widely used in engineering and natural sciences. The Euler method and improved Euler method can also be seen as special Runge--Kutta methods. When using the larger order $m$ in  Eq.~(\ref{eq:rk}), we need to compute iteratively a series of $K_i,i=1,2,... ,m$, increasing the computation cost for each step.

\begin{figure}[t]
    \centering
    \includegraphics[width=0.5\linewidth]{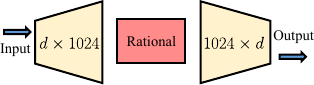}
    \caption{\textbf{Neural network structure in NeurVec.} NeurVec consists of two linear transformations layers (yellow) and one nonlinear activation function layer (red), which is a feed-forward neural network of one hidden layer ($L=1$) with the width $N_1=1024$. }
    \label{fig:nnarc}
\end{figure}

 \noindent\textbf{Implementation details of NeurVec}. We use a fully connected neural network to model  NeurVec in Eq.~(\ref{eq:iterative2}). The fully connected feed-forward neural network is the composition of $L$ nonlinear functions:
\begin{equation}\label{eqn:FNN}
	\phi(\mathbf{x};\bm{\theta}):=\mathbf{W_a} \circ \mathbf{h}_L \circ \mathbf{h}_{L-1} \circ \cdots \circ \mathbf{h}_{1}(\mathbf{x}),
\end{equation}
 where $\mathbf{h}_{\ell}(\mathbf{x})=\sigma\left(\mathbf{W}_\ell \mathbf{x} + \mathbf{b}_\ell \right)$ with $\mathbf{W}_\ell \in \mathbb{R}^{N_{\ell}\times N_{\ell-1}}$, $\mathbf{b}_\ell \in \mathbb{R}^{N_\ell}$ for $\ell=1,\dots,L$, $\mathbf{W_a}\in \mathbb{R}^{d\times N_L}$, $\mathbf{x}\in R^{d\times N}$, $\sigma$ is a nonlinear activation function, for example, a rectified linear unit (ReLU) $\sigma(x)=\max\{x,0\}$ or hyperbolic tangent function $\tanh(x)$, $d$ is the dimension of the state, and $N$ is the batch size. Each $\mathbf{h}_\ell$ is referred to as a hidden layer,  where  $N_\ell$ is the width of the $\ell$th layer. In this formulation, $\bm{\theta}:=\{\mathbf{W_a},\,\mathbf{W}_\ell,\,\mathbf{b}_\ell:1\leq \ell\leq L\}$ denotes the set of all parameters in $\phi$, which uniquely determines the underlying neural network.
In our implementation (Fig.~\ref{fig:nnarc}), the feed-forward neural network is of one hidden layer ($L=1$) with the width $N_1=1024$. The activation function used is the rational activation function\cite{boulle2020rational} defended by 
\begin{align}
\frac{a_3x^3+a_2x^2+a_1x^1+a_0}{b_2x^2+b_1x^1+b_0},
\end{align}
where $a_i,0\leq i\leq3$ and $b_i,0\leq i\leq2$ are initialized by constants $a_0=0.0218$, $a_1 = 0.5000$, $a_2 = 0.5957$, $a_3 = 1.1915$, $b_0 = 1.0000$, $b_1 = 0.0000$, $b_2 = 2.3830$, respectively. The parameters in $\mathbf{W}_a$ and $\mathbf{W}_\ell$ are initialized from  $\mathcal{U}[-1/\sqrt{N_0},1/\sqrt{N_0}]$ and $\mathcal{U}[-1/\sqrt{N_1},1/\sqrt{N_1}]$, respectively. We optimize the $\phi$ for 500 epochs with the  Adam optimizer. Moreover, we use the mean square error as the objective function in Eq.~(\ref{eq:objective}), and we set the initial learning rate to 1e-3.\\

\section*{Author contributions}
Z.Huang and S.Liang conceived and performed the experiments in this paper. H.Yang and L.Lin contributed materials and analysis tools. H.Zhang, Z.Huang and S.Liang analyzed the data and wrote the paper. All authors reviewed the manuscript.

\section*{Data and code availability}
The synthesized data for the \rv{different complex dynamical systems} and source codes for training and testing results are available at the online data warehouse: \url{https://github.com/dedekinds/NeurVec}. The source codes are released under MIT license.

\section*{Acknowledgement}
 Z. Huang and L. Lin was supported in part by  National Key R\&D Program of China under Grant No. 2020AAA0109700, National Natural Science Foundation of China (NSFC) under Grant No.61836012, Guangdong Basic and Applied Basic Research Foundation No. 2023A1515011374, National Natural Science Foundation of China (NSFC) under Grant No. 62206314, GuangDong Basic and Applied Basic Research Foundation under Grant No. 2022A1515011835. H. Y. was partially supported by the US National Science Foundation under award DMS-2244988, DMS-2206333, and the Office of Naval Research Award N00014-23-1-2007. Argonne National Laboratory's work was supported by the U.S. Department of Energy, Office of Science, Office of Advanced Scientific Computing Research, Scientific Discovery through Advanced Computing (SciDAC) program through the FASTMath Institute under contract DE-AC02-06CH11357 at Argonne National
Laboratory. We also acknowledge funding support from ASCR for DOE-FOA-2493, ``Data-intensive scientific machine learning''.

\bibliography{sample}

\begin{thebibliography}{10}
\urlstyle{rm}
\expandafter\ifx\csname url\endcsname\relax
  \def\url#1{\texttt{#1}}\fi
\expandafter\ifx\csname urlprefix\endcsname\relax\def\urlprefix{URL }\fi
\expandafter\ifx\csname doiprefix\endcsname\relax\def\doiprefix{DOI: }\fi
\providecommand{\bibinfo}[2]{#2}
\providecommand{\eprint}[2][]{\url{#2}}

\bibitem{bottcher2022ai}
\bibinfo{author}{Böttcher, L.}, \bibinfo{author}{Antulov-Fantulin, N.} \& \bibinfo{author}{Asikis, T.}
\newblock \bibinfo{journal}{\bibinfo{title}{{AI Pontryagin} or how artificial neural networks learn to control dynamical systems}}.
\newblock {\emph{\JournalTitle{Nature Communications}}} \textbf{\bibinfo{volume}{13}}, \bibinfo{pages}{1--9} (\bibinfo{year}{2022}).

\bibitem{stuart1998dynamical}
\bibinfo{author}{Stuart, A.} \& \bibinfo{author}{Humphries, A.~R.}
\newblock \emph{\bibinfo{title}{Dynamical systems and numerical analysis}}, vol.~\bibinfo{volume}{2} (\bibinfo{publisher}{Cambridge University Press}, \bibinfo{year}{1998}).

\bibitem{harlim2021machine}
\bibinfo{author}{Harlim, J.}, \bibinfo{author}{Jiang, S.~W.}, \bibinfo{author}{Liang, S.} \& \bibinfo{author}{Yang, H.}
\newblock \bibinfo{journal}{\bibinfo{title}{Machine learning for prediction with missing dynamics}}.
\newblock {\emph{\JournalTitle{Journal of Computational Physics}}} \textbf{\bibinfo{volume}{428}}, \bibinfo{pages}{109922} (\bibinfo{year}{2021}).

\bibitem{kou2019nitrogen}
\bibinfo{author}{Kou-Giesbrecht, S.} \& \bibinfo{author}{Menge, D.}
\newblock \bibinfo{journal}{\bibinfo{title}{Nitrogen-fixing trees could exacerbate climate change under elevated nitrogen deposition}}.
\newblock {\emph{\JournalTitle{Nature Communications}}} \textbf{\bibinfo{volume}{10}}, \bibinfo{pages}{1--8} (\bibinfo{year}{2019}).

\bibitem{benn2019general}
\bibinfo{author}{Benn, D.}, \bibinfo{author}{Fowler, A.~C.}, \bibinfo{author}{Hewitt, I.} \& \bibinfo{author}{Sevestre, H.}
\newblock \bibinfo{journal}{\bibinfo{title}{A general theory of glacier surges}}.
\newblock {\emph{\JournalTitle{Journal of Glaciology}}} \textbf{\bibinfo{volume}{65}}, \bibinfo{pages}{701--716} (\bibinfo{year}{2019}).

\bibitem{owoyele2022chemnode}
\bibinfo{author}{Owoyele, O.} \& \bibinfo{author}{Pal, P.}
\newblock \bibinfo{journal}{\bibinfo{title}{Chemnode: A neural ordinary differential equations framework for efficient chemical kinetic solvers}}.
\newblock {\emph{\JournalTitle{Energy and AI}}} \textbf{\bibinfo{volume}{7}}, \bibinfo{pages}{100118} (\bibinfo{year}{2022}).

\bibitem{zhang2011chemical}
\bibinfo{author}{Zhang, H.}, \bibinfo{author}{Linford, J.~C.}, \bibinfo{author}{Sandu, A.} \& \bibinfo{author}{Sander, R.}
\newblock \bibinfo{journal}{\bibinfo{title}{Chemical mechanism solvers in air quality models}}.
\newblock {\emph{\JournalTitle{Atmosphere}}} \textbf{\bibinfo{volume}{2}}, \bibinfo{pages}{510--532} (\bibinfo{year}{2011}).

\bibitem{delpini2013evolution}
\bibinfo{author}{Delpini, D.} \emph{et~al.}
\newblock \bibinfo{journal}{\bibinfo{title}{Evolution of controllability in interbank networks}}.
\newblock {\emph{\JournalTitle{Scientific Reports}}} \textbf{\bibinfo{volume}{3}}, \bibinfo{pages}{1--5} (\bibinfo{year}{2013}).

\bibitem{gholami2022impact}
\bibinfo{author}{Gholami, A.} \& \bibinfo{author}{Sun, X.~A.}
\newblock \bibinfo{journal}{\bibinfo{title}{The impact of damping in second-order dynamical systems with applications to power grid stability}}.
\newblock {\emph{\JournalTitle{SIAM Journal on Applied Dynamical Systems}}} \textbf{\bibinfo{volume}{21}}, \bibinfo{pages}{405--437} (\bibinfo{year}{2022}).

\bibitem{schafer2018dynamically}
\bibinfo{author}{Sch{\"a}fer, B.}, \bibinfo{author}{Witthaut, D.}, \bibinfo{author}{Timme, M.} \& \bibinfo{author}{Latora, V.}
\newblock \bibinfo{journal}{\bibinfo{title}{Dynamically induced cascading failures in power grids}}.
\newblock {\emph{\JournalTitle{Nature Communications}}} \textbf{\bibinfo{volume}{9}}, \bibinfo{pages}{1--13} (\bibinfo{year}{2018}).

\bibitem{saberi2020simple}
\bibinfo{author}{Saberi, M.} \emph{et~al.}
\newblock \bibinfo{journal}{\bibinfo{title}{A simple contagion process describes spreading of traffic jams in urban networks}}.
\newblock {\emph{\JournalTitle{Nature Communications}}} \textbf{\bibinfo{volume}{11}}, \bibinfo{pages}{1--9} (\bibinfo{year}{2020}).

\bibitem{fan2020network}
\bibinfo{author}{Fan, C.}, \bibinfo{author}{Jiang, X.} \& \bibinfo{author}{Mostafavi, A.}
\newblock \bibinfo{journal}{\bibinfo{title}{A network percolation-based contagion model of flood propagation and recession in urban road networks}}.
\newblock {\emph{\JournalTitle{Scientific Reports}}} \textbf{\bibinfo{volume}{10}}, \bibinfo{pages}{1--12} (\bibinfo{year}{2020}).

\bibitem{aulin2021design}
\bibinfo{author}{Aulin, L.}, \bibinfo{author}{Liakopoulos, A.}, \bibinfo{author}{van~der Graaf, P.~H.}, \bibinfo{author}{Rozen, D.~E.} \& \bibinfo{author}{van Hasselt, J.}
\newblock \bibinfo{journal}{\bibinfo{title}{Design principles of collateral sensitivity-based dosing strategies}}.
\newblock {\emph{\JournalTitle{Nature Communications}}} \textbf{\bibinfo{volume}{12}}, \bibinfo{pages}{1--14} (\bibinfo{year}{2021}).

\bibitem{butner2021mathematical}
\bibinfo{author}{Butner, J.~D.} \emph{et~al.}
\newblock \bibinfo{journal}{\bibinfo{title}{A mathematical model for the quantification of a patient’s sensitivity to checkpoint inhibitors and long-term tumour burden}}.
\newblock {\emph{\JournalTitle{Nature Biomedical Engineering}}} \textbf{\bibinfo{volume}{5}}, \bibinfo{pages}{297--308} (\bibinfo{year}{2021}).

\bibitem{wicha2017general}
\bibinfo{author}{Wicha, S.~G.}, \bibinfo{author}{Chen, C.}, \bibinfo{author}{Clewe, O.} \& \bibinfo{author}{Simonsson, U.~S.}
\newblock \bibinfo{journal}{\bibinfo{title}{A general pharmacodynamic interaction model identifies perpetrators and victims in drug interactions}}.
\newblock {\emph{\JournalTitle{Nature communications}}} \textbf{\bibinfo{volume}{8}}, \bibinfo{pages}{1--11} (\bibinfo{year}{2017}).

\bibitem{butcher2016numerical}
\bibinfo{author}{Butcher, J.~C.}
\newblock \emph{\bibinfo{title}{Numerical methods for ordinary differential equations}} (\bibinfo{publisher}{John Wiley and Sons}, \bibinfo{year}{2016}).

\bibitem{ames2014numerical}
\bibinfo{author}{Ames, W.~F.}
\newblock \emph{\bibinfo{title}{Numerical methods for partial differential equations}} (\bibinfo{publisher}{Academic press}, \bibinfo{year}{2014}).

\bibitem{shampine2018numerical}
\bibinfo{author}{Shampine, L.~F.}
\newblock \emph{\bibinfo{title}{Numerical solution of ordinary differential equations}} (\bibinfo{publisher}{Routledge}, \bibinfo{year}{2018}).

\bibitem{figueiras2019qmblender}
\bibinfo{author}{Figueiras, E.}, \bibinfo{author}{Olivieri, D.}, \bibinfo{author}{Paredes, A.} \& \bibinfo{author}{Michinel, H.}
\newblock \bibinfo{journal}{\bibinfo{title}{{QMBlender: Particle-based visualization of 3D quantum wave function dynamics}}}.
\newblock {\emph{\JournalTitle{Journal of Computational Science}}} \textbf{\bibinfo{volume}{35}}, \bibinfo{pages}{44--56} (\bibinfo{year}{2019}).

\bibitem{xi2019survey}
\bibinfo{author}{Xi, R.} \emph{et~al.}
\newblock \bibinfo{journal}{\bibinfo{title}{Survey on smoothed particle hydrodynamics and the particle systems}}.
\newblock {\emph{\JournalTitle{IEEE Access}}} \textbf{\bibinfo{volume}{8}}, \bibinfo{pages}{3087--3105} (\bibinfo{year}{2019}).

\bibitem{zhang2020cumuliform}
\bibinfo{author}{Zhang, Z.}, \bibinfo{author}{Zhang, Y.}, \bibinfo{author}{Li, Y.} \& \bibinfo{author}{Liang, X.}
\newblock \bibinfo{title}{Cumuliform cloud animation control based on natural images}.
\newblock In \emph{\bibinfo{booktitle}{2020 International Conference on Virtual Reality and Visualization (ICVRV)}}, \bibinfo{pages}{218--224} (\bibinfo{organization}{IEEE}, \bibinfo{year}{2020}).

\bibitem{scher2021ensemble}
\bibinfo{author}{Scher, S.} \& \bibinfo{author}{Messori, G.}
\newblock \bibinfo{journal}{\bibinfo{title}{Ensemble methods for neural network-based weather forecasts}}.
\newblock {\emph{\JournalTitle{Journal of Advances in Modeling Earth Systems}}} \textbf{\bibinfo{volume}{13}} (\bibinfo{year}{2021}).

\bibitem{hu2021characteristics}
\bibinfo{author}{Hu, B.}, \bibinfo{author}{Guo, H.}, \bibinfo{author}{Zhou, P.} \& \bibinfo{author}{Shi, Z.-L.}
\newblock \bibinfo{journal}{\bibinfo{title}{{Characteristics of SARS-CoV-2 and COVID-19}}}.
\newblock {\emph{\JournalTitle{Nature Reviews Microbiology}}} \textbf{\bibinfo{volume}{19}}, \bibinfo{pages}{141--154} (\bibinfo{year}{2021}).

\bibitem{bousquet2022deep}
\bibinfo{author}{Bousquet, A.}, \bibinfo{author}{Conrad, W.~H.}, \bibinfo{author}{Sadat, S.~O.}, \bibinfo{author}{Vardanyan, N.} \& \bibinfo{author}{Hong, Y.}
\newblock \bibinfo{journal}{\bibinfo{title}{Deep learning forecasting using time-varying parameters of the {SIRD model for Covid-19}}}.
\newblock {\emph{\JournalTitle{Scientific Reports}}} \textbf{\bibinfo{volume}{12}}, \bibinfo{pages}{1--13} (\bibinfo{year}{2022}).

\bibitem{beira2021differential}
\bibinfo{author}{Beira, M.~J.} \& \bibinfo{author}{Sebasti{\~a}o, P.~J.}
\newblock \bibinfo{journal}{\bibinfo{title}{A differential equations model-fitting analysis of {COVID-19} epidemiological data to explain multi-wave dynamics}}.
\newblock {\emph{\JournalTitle{Scientific Reports}}} \textbf{\bibinfo{volume}{11}}, \bibinfo{pages}{1--13} (\bibinfo{year}{2021}).

\bibitem{choi2021optimal}
\bibinfo{author}{Choi, W.} \& \bibinfo{author}{Shim, E.}
\newblock \bibinfo{journal}{\bibinfo{title}{Optimal strategies for social distancing and testing to control {COVID-19}}}.
\newblock {\emph{\JournalTitle{Journal of Theoretical Biology}}} \textbf{\bibinfo{volume}{512}}, \bibinfo{pages}{110568} (\bibinfo{year}{2021}).

\bibitem{hsiang2020effect}
\bibinfo{author}{Hsiang, S.} \emph{et~al.}
\newblock \bibinfo{journal}{\bibinfo{title}{The effect of large-scale anti-contagion policies on the {COVID-19} pandemic}}.
\newblock {\emph{\JournalTitle{Nature}}} \textbf{\bibinfo{volume}{584}}, \bibinfo{pages}{262--267} (\bibinfo{year}{2020}).

\bibitem{tregoning2021progress}
\bibinfo{author}{Tregoning, J.~S.}, \bibinfo{author}{Flight, K.~E.}, \bibinfo{author}{Higham, S.~L.}, \bibinfo{author}{Wang, Z.} \& \bibinfo{author}{Pierce, B.~F.}
\newblock \bibinfo{journal}{\bibinfo{title}{Progress of the {COVID-19} vaccine effort: viruses, vaccines and variants versus efficacy, effectiveness and escape}}.
\newblock {\emph{\JournalTitle{Nature Reviews Immunology}}} \textbf{\bibinfo{volume}{21}}, \bibinfo{pages}{626--636} (\bibinfo{year}{2021}).

\bibitem{yuan2019wave}
\bibinfo{author}{Yuan, F.}, \bibinfo{author}{Zhang, L.}, \bibinfo{author}{Xia, X.}, \bibinfo{author}{Huang, Q.} \& \bibinfo{author}{Li, X.}
\newblock \bibinfo{journal}{\bibinfo{title}{A wave-shaped deep neural network for smoke density estimation}}.
\newblock {\emph{\JournalTitle{IEEE Transactions on Image Processing}}} \textbf{\bibinfo{volume}{29}}, \bibinfo{pages}{2301--2313} (\bibinfo{year}{2019}).

\bibitem{tumanov2021data}
\bibinfo{author}{Tumanov, E.}, \bibinfo{author}{Korobchenko, D.} \& \bibinfo{author}{Chentanez, N.}
\newblock \bibinfo{journal}{\bibinfo{title}{Data-driven particle-based liquid simulation with deep learning utilizing sub-pixel convolution}}.
\newblock {\emph{\JournalTitle{Proceedings of the ACM on Computer Graphics and Interactive Techniques}}} \textbf{\bibinfo{volume}{4}}, \bibinfo{pages}{1--16} (\bibinfo{year}{2021}).

\bibitem{kolb2004hardware}
\bibinfo{author}{Kolb, A.}, \bibinfo{author}{Latta, L.} \& \bibinfo{author}{Rezk-Salama, C.}
\newblock \bibinfo{title}{Hardware-based simulation and collision detection for large particle systems}.
\newblock In \emph{\bibinfo{booktitle}{Proceedings of the ACM SIGGRAPH/EUROGRAPHICS conference on Graphics hardware}}, \bibinfo{pages}{123--131} (\bibinfo{year}{2004}).

\bibitem{luo2022using}
\bibinfo{author}{Luo, H.} \& \bibinfo{author}{Wu, Y.}
\newblock \bibinfo{journal}{\bibinfo{title}{Using virtual reality technology to construct computer-aided animation material development}}.
\newblock {\emph{\JournalTitle{Computer-Aided Design \& Applications}}} \textbf{\bibinfo{volume}{19}}, \bibinfo{pages}{155--166} (\bibinfo{year}{2022}).

\bibitem{bellprat2019towards}
\bibinfo{author}{Bellprat, O.}, \bibinfo{author}{Guemas, V.}, \bibinfo{author}{Doblas-Reyes, F.} \& \bibinfo{author}{Donat, M.~G.}
\newblock \bibinfo{journal}{\bibinfo{title}{Towards reliable extreme weather and climate event attribution}}.
\newblock {\emph{\JournalTitle{Nature Communications}}} \textbf{\bibinfo{volume}{10}}, \bibinfo{pages}{1--7} (\bibinfo{year}{2019}).

\bibitem{touma2021human}
\bibinfo{author}{Touma, D.}, \bibinfo{author}{Stevenson, S.}, \bibinfo{author}{Lehner, F.} \& \bibinfo{author}{Coats, S.}
\newblock \bibinfo{journal}{\bibinfo{title}{Human-driven greenhouse gas and aerosol emissions cause distinct regional impacts on extreme fire weather}}.
\newblock {\emph{\JournalTitle{Nature Communications}}} \textbf{\bibinfo{volume}{12}}, \bibinfo{pages}{1--8} (\bibinfo{year}{2021}).

\bibitem{palmer2013singular}
\bibinfo{author}{Palmer, T.} \& \bibinfo{author}{Zanna, L.}
\newblock \bibinfo{journal}{\bibinfo{title}{Singular vectors, predictability and ensemble forecasting for weather and climate}}.
\newblock {\emph{\JournalTitle{Journal of Physics A: Mathematical and Theoretical}}} \textbf{\bibinfo{volume}{46}}, \bibinfo{pages}{254018} (\bibinfo{year}{2013}).

\bibitem{wu2021ensemble}
\bibinfo{author}{Wu, H.} \& \bibinfo{author}{Levinson, D.}
\newblock \bibinfo{journal}{\bibinfo{title}{The ensemble approach to forecasting: a review and synthesis}}.
\newblock {\emph{\JournalTitle{Transportation Research Part C: Emerging Technologies}}} \textbf{\bibinfo{volume}{132}}, \bibinfo{pages}{103357} (\bibinfo{year}{2021}).

\bibitem{popov2021multifidelity}
\bibinfo{author}{Popov, A.~A.}, \bibinfo{author}{Mou, C.}, \bibinfo{author}{Sandu, A.} \& \bibinfo{author}{Iliescu, T.}
\newblock \bibinfo{journal}{\bibinfo{title}{A multifidelity ensemble kalman filter with reduced order control variates}}.
\newblock {\emph{\JournalTitle{SIAM Journal on Scientific Computing}}} \textbf{\bibinfo{volume}{43}}, \bibinfo{pages}{A1134--A1162} (\bibinfo{year}{2021}).

\bibitem{brodtkorb2013graphics}
\bibinfo{author}{Brodtkorb, A.~R.}, \bibinfo{author}{Hagen, T.~R.} \& \bibinfo{author}{Sætra, M.~L.}
\newblock \bibinfo{journal}{\bibinfo{title}{Graphics processing unit {(GPU)} programming strategies and trends in {GPU} computing}}.
\newblock {\emph{\JournalTitle{Journal of Parallel and Distributed Computing}}} \textbf{\bibinfo{volume}{73}}, \bibinfo{pages}{4--13} (\bibinfo{year}{2013}).

\bibitem{jouppi2017datacenter}
\bibinfo{author}{Jouppi, N.~P.} \emph{et~al.}
\newblock \bibinfo{title}{In-datacenter performance analysis of a tensor processing unit}.
\newblock In \emph{\bibinfo{booktitle}{Proceedings of the 44th annual international symposium on computer architecture}}, \bibinfo{pages}{1--12} (\bibinfo{year}{2017}).

\bibitem{tan2020fastva}
\bibinfo{author}{Tan, T.} \& \bibinfo{author}{Cao, G.}
\newblock \bibinfo{title}{{FastVA:} deep learning video analytics through edge processing and {NPU} in mobile}.
\newblock In \emph{\bibinfo{booktitle}{IEEE INFOCOM 2020-IEEE Conference on Computer Communications}}, \bibinfo{pages}{1947--1956} (\bibinfo{organization}{IEEE}, \bibinfo{year}{2020}).

\bibitem{liao2002high}
\bibinfo{author}{Liao, Y.} \& \bibinfo{author}{Roberts, D.~B.}
\newblock \bibinfo{journal}{\bibinfo{title}{A high-performance and low-power 32-bit multiply-accumulate unit with single-instruction-multiple-data {(SIMD)} feature}}.
\newblock {\emph{\JournalTitle{IEEE Journal of Solid-State Circuits}}} \textbf{\bibinfo{volume}{37}}, \bibinfo{pages}{926--931} (\bibinfo{year}{2002}).

\bibitem{fehlberg1969low}
\bibinfo{author}{Fehlberg, E.}
\newblock \emph{\bibinfo{title}{{Low-order classical {Runge--Kutta} formulas with stepsize control and their application to some heat transfer problems}}}, vol. \bibinfo{volume}{315} (\bibinfo{publisher}{National Aeronautics and Space Administration}, \bibinfo{year}{1969}).

\bibitem{chen2018neural}
\bibinfo{author}{Chen, R.~T.}, \bibinfo{author}{Rubanova, Y.}, \bibinfo{author}{Bettencourt, J.} \& \bibinfo{author}{Duvenaud, D.~K.}
\newblock \bibinfo{journal}{\bibinfo{title}{Neural ordinary differential equations}}.
\newblock {\emph{\JournalTitle{Advances in Neural Information Processing Systems}}} \textbf{\bibinfo{volume}{31}} (\bibinfo{year}{2018}).

\bibitem{liang2022stiffnessaware}
\bibinfo{author}{Liang, S.}, \bibinfo{author}{Huang, Z.} \& \bibinfo{author}{Zhang, H.}
\newblock \bibinfo{title}{Stiffness-aware neural network for learning {Hamiltonian} systems}.
\newblock In \emph{\bibinfo{booktitle}{International Conference on Learning Representations}} (\bibinfo{year}{2022}).

\bibitem{li2021fourier}
\bibinfo{author}{Li, Z.} \emph{et~al.}
\newblock \bibinfo{title}{Fourier neural operator for parametric partial differential equations}.
\newblock In \emph{\bibinfo{booktitle}{International Conference on Learning Representations}} (\bibinfo{year}{2021}).

\bibitem{pan2018long}
\bibinfo{author}{Pan, S.} \& \bibinfo{author}{Duraisamy, K.}
\newblock \bibinfo{journal}{\bibinfo{title}{Long-time predictive modeling of nonlinear dynamical systems using neural networks}}.
\newblock {\emph{\JournalTitle{Complexity}}} \textbf{\bibinfo{volume}{2018}}, \bibinfo{pages}{1--26} (\bibinfo{year}{2018}).

\bibitem{liu2022predicting}
\bibinfo{author}{Liu, X.-Y.}, \bibinfo{author}{Sun, H.}, \bibinfo{author}{Zhu, M.}, \bibinfo{author}{Lu, L.} \& \bibinfo{author}{Wang, J.-X.}
\newblock \bibinfo{journal}{\bibinfo{title}{Predicting parametric spatiotemporal dynamics by multi-resolution pde structure-preserved deep learning}}.
\newblock {\emph{\JournalTitle{arXiv preprint arXiv:2205.03990}}}  (\bibinfo{year}{2022}).

\bibitem{chen2021generalized}
\bibinfo{author}{Chen, Z.} \& \bibinfo{author}{Xiu, D.}
\newblock \bibinfo{journal}{\bibinfo{title}{On generalized residual network for deep learning of unknown dynamical systems}}.
\newblock {\emph{\JournalTitle{Journal of Computational Physics}}} \textbf{\bibinfo{volume}{438}}, \bibinfo{pages}{110362} (\bibinfo{year}{2021}).

\bibitem{dresdner2022learning}
\bibinfo{author}{Dresdner, G.} \emph{et~al.}
\newblock \bibinfo{journal}{\bibinfo{title}{Learning to correct spectral methods for simulating turbulent flows}}.
\newblock {\emph{\JournalTitle{arXiv preprint arXiv:2207.00556}}}  (\bibinfo{year}{2022}).

\bibitem{lu2021learning}
\bibinfo{author}{Lu, L.}, \bibinfo{author}{Jin, P.}, \bibinfo{author}{Pang, G.}, \bibinfo{author}{Zhang, Z.} \& \bibinfo{author}{Karniadakis, G.~E.}
\newblock \bibinfo{journal}{\bibinfo{title}{Learning nonlinear operators via {DeepONet} based on the universal approximation theorem of operators}}.
\newblock {\emph{\JournalTitle{Nature Machine Intelligence}}} \textbf{\bibinfo{volume}{3}}, \bibinfo{pages}{218--229} (\bibinfo{year}{2021}).

\bibitem{choudhary2020physics}
\bibinfo{author}{Choudhary, A.} \emph{et~al.}
\newblock \bibinfo{journal}{\bibinfo{title}{Physics-enhanced neural networks learn order and chaos}}.
\newblock {\emph{\JournalTitle{Physical Review E}}} \textbf{\bibinfo{volume}{101}}, \bibinfo{pages}{062207} (\bibinfo{year}{2020}).

\bibitem{han2021adaptable}
\bibinfo{author}{Han, C.-D.}, \bibinfo{author}{Glaz, B.}, \bibinfo{author}{Haile, M.} \& \bibinfo{author}{Lai, Y.-C.}
\newblock \bibinfo{journal}{\bibinfo{title}{Adaptable {Hamiltonian} neural networks}}.
\newblock {\emph{\JournalTitle{Physical Review Research}}} \textbf{\bibinfo{volume}{3}}, \bibinfo{pages}{023156} (\bibinfo{year}{2021}).

\bibitem{greydanus2019hamiltonian}
\bibinfo{author}{Greydanus, S.}, \bibinfo{author}{Dzamba, M.} \& \bibinfo{author}{Yosinski, J.}
\newblock \bibinfo{journal}{\bibinfo{title}{Hamiltonian neural networks}}.
\newblock {\emph{\JournalTitle{Advances in Neural Information Processing Systems}}} \textbf{\bibinfo{volume}{32}} (\bibinfo{year}{2019}).

\bibitem{huang1997adaptive}
\bibinfo{author}{Huang, W.} \& \bibinfo{author}{Leimkuhler, B.}
\newblock \bibinfo{journal}{\bibinfo{title}{The adaptive {Verlet} method}}.
\newblock {\emph{\JournalTitle{SIAM Journal on Scientific Computing}}} \textbf{\bibinfo{volume}{18}}, \bibinfo{pages}{239--256} (\bibinfo{year}{1997}).

\bibitem{wang2022stable}
\bibinfo{author}{Wang, X.}, \bibinfo{author}{Han, Y.}, \bibinfo{author}{Xue, W.}, \bibinfo{author}{Yang, G.} \& \bibinfo{author}{Zhang, G.~J.}
\newblock \bibinfo{journal}{\bibinfo{title}{Stable climate simulations using a realistic general circulation model with neural network parameterizations for atmospheric moist physics and radiation processes}}.
\newblock {\emph{\JournalTitle{Geoscientific Model Development}}} \textbf{\bibinfo{volume}{15}}, \bibinfo{pages}{3923--3940} (\bibinfo{year}{2022}).

\bibitem{paszke2019pytorch}
\bibinfo{author}{Paszke, A.} \emph{et~al.}
\newblock \bibinfo{journal}{\bibinfo{title}{Pytorch: An imperative style, high-performance deep learning library}}.
\newblock {\emph{\JournalTitle{Advances in neural information processing systems}}} \textbf{\bibinfo{volume}{32}} (\bibinfo{year}{2019}).

\bibitem{abadi2016tensorflow}
\bibinfo{author}{Abadi, M.} \emph{et~al.}
\newblock \bibinfo{title}{Tensorflow: A system for large-scale machine learning}.
\newblock In \emph{\bibinfo{booktitle}{12th USENIX symposium on operating systems design and implementation (OSDI 16)}}, \bibinfo{pages}{265--283} (\bibinfo{year}{2016}).

\bibitem{boulle2020rational}
\bibinfo{author}{Boullé, N.}, \bibinfo{author}{Nakatsukasa, Y.} \& \bibinfo{author}{Townsend, A.}
\newblock \bibinfo{journal}{\bibinfo{title}{Rational neural networks}}.
\newblock {\emph{\JournalTitle{Advances in Neural Information Processing Systems}}} \textbf{\bibinfo{volume}{33}}, \bibinfo{pages}{14243--14253} (\bibinfo{year}{2020}).

\bibitem{Chen2020Symplectic}
\bibinfo{author}{Chen, Z.}, \bibinfo{author}{Zhang, J.}, \bibinfo{author}{Arjovsky, M.} \& \bibinfo{author}{Bottou, L.}
\newblock \bibinfo{title}{Symplectic recurrent neural networks}.
\newblock In \emph{\bibinfo{booktitle}{International Conference on Learning Representations}} (\bibinfo{year}{2020}).

\bibitem{feit1984wave}
\bibinfo{author}{Feit, M.} \& \bibinfo{author}{Fleck~Jr, J.}
\newblock \bibinfo{journal}{\bibinfo{title}{Wave packet dynamics and chaos in the hénnon--heiles system}}.
\newblock {\emph{\JournalTitle{The Journal of Chemical Physics}}} \textbf{\bibinfo{volume}{80}}, \bibinfo{pages}{2578--2584} (\bibinfo{year}{1984}).

\bibitem{NEURIPS2020_439fca36}
\bibinfo{author}{DiPietro, D.}, \bibinfo{author}{Xiong, S.} \& \bibinfo{author}{Zhu, B.}
\newblock \bibinfo{title}{Sparse symplectically integrated neural networks}.
\newblock In \bibinfo{editor}{Larochelle, H.}, \bibinfo{editor}{Ranzato, M.}, \bibinfo{editor}{Hadsell, R.}, \bibinfo{editor}{Balcan, M.} \& \bibinfo{editor}{Lin, H.} (eds.) \emph{\bibinfo{booktitle}{Advances in Neural Information Processing Systems}}, vol.~\bibinfo{volume}{33}, \bibinfo{pages}{6074--6085} (\bibinfo{publisher}{Curran Associates, Inc.}, \bibinfo{year}{2020}).

\bibitem{breitenberger1981elastic}
\bibinfo{author}{Breitenberger, E.} \& \bibinfo{author}{Mueller, R.~D.}
\newblock \bibinfo{journal}{\bibinfo{title}{The elastic pendulum: a nonlinear paradigm}}.
\newblock {\emph{\JournalTitle{Journal of Mathematical Physics}}} \textbf{\bibinfo{volume}{22}}, \bibinfo{pages}{1196--1210} (\bibinfo{year}{1981}).

\bibitem{lopes2017dynamics}
\bibinfo{author}{Lopes, A.~M.} \& \bibinfo{author}{Tenreiro~Machado, J.}
\newblock \bibinfo{journal}{\bibinfo{title}{Dynamics of the {N-link} pendulum: a fractional perspective}}.
\newblock {\emph{\JournalTitle{International Journal of Control}}} \textbf{\bibinfo{volume}{90}}, \bibinfo{pages}{1192--1200} (\bibinfo{year}{2017}).

\bibitem{atkinson2011numerical}
\bibinfo{author}{Atkinson, K.}, \bibinfo{author}{Han, W.} \& \bibinfo{author}{Stewart, D.~E.}
\newblock \emph{\bibinfo{title}{Numerical solution of ordinary differential equations}} (\bibinfo{publisher}{John Wiley and Sons}, \bibinfo{year}{2011}).

\bibitem{du2018gradient}
\bibinfo{author}{Du, S.~S.}, \bibinfo{author}{Zhai, X.}, \bibinfo{author}{Poczos, B.} \& \bibinfo{author}{Singh, A.}
\newblock \bibinfo{title}{Gradient descent provably optimizes over-parameterized neural networks}.
\newblock In \emph{\bibinfo{booktitle}{International Conference on Learning Representations}} (\bibinfo{year}{2019}).

\bibitem{jacot2018neural}
\bibinfo{author}{Jacot, A.}, \bibinfo{author}{Gabriel, F.} \& \bibinfo{author}{Hongler, C.}
\newblock \bibinfo{journal}{\bibinfo{title}{Neural tangent kernel: Convergence and generalization in neural networks}}.
\newblock {\emph{\JournalTitle{Advances in Neural Information Processing Systems}}} \textbf{\bibinfo{volume}{31}} (\bibinfo{year}{2018}).

\bibitem{chizat2019lazy}
\bibinfo{author}{Chizat, L.}, \bibinfo{author}{Oyallon, E.} \& \bibinfo{author}{Bach, F.}
\newblock \bibinfo{journal}{\bibinfo{title}{On lazy training in differentiable programming}}.
\newblock {\emph{\JournalTitle{Advances in Neural Information Processing Systems}}} \textbf{\bibinfo{volume}{32}} (\bibinfo{year}{2019}).

\bibitem{lu2017expressive}
\bibinfo{author}{Lu, Z.}, \bibinfo{author}{Pu, H.}, \bibinfo{author}{Wang, F.}, \bibinfo{author}{Hu, Z.} \& \bibinfo{author}{Wang, L.}
\newblock \bibinfo{journal}{\bibinfo{title}{The expressive power of neural networks: A view from the width}}.
\newblock {\emph{\JournalTitle{Advances in neural information processing systems}}} \textbf{\bibinfo{volume}{30}} (\bibinfo{year}{2017}).

\bibitem{hu2018squeeze}
\bibinfo{author}{Hu, J.}, \bibinfo{author}{Shen, L.} \& \bibinfo{author}{Sun, G.}
\newblock \bibinfo{title}{Squeeze-and-excitation networks}.
\newblock In \emph{\bibinfo{booktitle}{Proceedings of the IEEE Conference on Computer Vision and Pattern Recognition}}, \bibinfo{pages}{7132--7141} (\bibinfo{year}{2018}).

\bibitem{huang2020dianet}
\bibinfo{author}{Huang, Z.}, \bibinfo{author}{Liang, S.}, \bibinfo{author}{Liang, M.} \& \bibinfo{author}{Yang, H.}
\newblock \bibinfo{title}{Dianet: Dense-and-implicit attention network}.
\newblock In \emph{\bibinfo{booktitle}{Proceedings of the AAAI Conference on Artificial Intelligence}}, vol.~\bibinfo{volume}{34}, \bibinfo{pages}{4206--4214} (\bibinfo{year}{2020}).

\bibitem{liang2020instance}
\bibinfo{author}{Liang, S.}, \bibinfo{author}{Huang, Z.}, \bibinfo{author}{Liang, M.} \& \bibinfo{author}{Yang, H.}
\newblock \bibinfo{title}{Instance enhancement batch normalization: An adaptive regulator of batch noise}.
\newblock In \emph{\bibinfo{booktitle}{Proceedings of the AAAI Conference on Artificial Intelligence}}, vol.~\bibinfo{volume}{34}, \bibinfo{pages}{4819--4827} (\bibinfo{year}{2020}).

\bibitem{he2021blending}
\bibinfo{author}{He, W.}, \bibinfo{author}{Huang, Z.}, \bibinfo{author}{Liang, M.}, \bibinfo{author}{Liang, S.} \& \bibinfo{author}{Yang, H.}
\newblock \bibinfo{title}{Blending pruning criteria for convolutional neural networks}.
\newblock In \emph{\bibinfo{booktitle}{International Conference on Artificial Neural Networks}}, \bibinfo{pages}{3--15} (\bibinfo{organization}{Springer}, \bibinfo{year}{2021}).

\bibitem{liu2018darts}
\bibinfo{author}{Liu, H.}, \bibinfo{author}{Simonyan, K.} \& \bibinfo{author}{Yang, Y.}
\newblock \bibinfo{title}{Darts: Differentiable architecture search}.
\newblock In \emph{\bibinfo{booktitle}{International Conference on Learning Representations}} (\bibinfo{year}{2018}).

\bibitem{huang2021rethinking}
\bibinfo{author}{Huang, Z.}, \bibinfo{author}{Shao, W.}, \bibinfo{author}{Wang, X.}, \bibinfo{author}{Lin, L.} \& \bibinfo{author}{Luo, P.}
\newblock \bibinfo{journal}{\bibinfo{title}{Rethinking the pruning criteria for convolutional neural network}}.
\newblock {\emph{\JournalTitle{Advances in Neural Information Processing Systems}}} \textbf{\bibinfo{volume}{34}}, \bibinfo{pages}{16305--16318} (\bibinfo{year}{2021}).

\bibitem{kenton2019bert}
\bibinfo{author}{Kenton, J. D. M.-W.~C.} \& \bibinfo{author}{Toutanova, L.~K.}
\newblock \bibinfo{title}{Bert: Pre-training of deep bidirectional transformers for language understanding}.
\newblock In \emph{\bibinfo{booktitle}{Proceedings of NAACL-HLT}}, \bibinfo{pages}{4171--4186} (\bibinfo{year}{2019}).

\bibitem{wu2019large}
\bibinfo{author}{Wu, Y.} \emph{et~al.}
\newblock \bibinfo{title}{Large scale incremental learning}.
\newblock In \emph{\bibinfo{booktitle}{Proceedings of the IEEE/CVF Conference on Computer Vision and Pattern Recognition}}, \bibinfo{pages}{374--382} (\bibinfo{year}{2019}).

\bibitem{huang2023robust}
\bibinfo{author}{Huang, Z.}, \bibinfo{author}{Liang, M.} \& \bibinfo{author}{Lin, L.}
\newblock \bibinfo{journal}{\bibinfo{title}{On robust numerical solver for ode via self-attention mechanism}}.
\newblock {\emph{\JournalTitle{arXiv preprint arXiv:2302.10184}}}  (\bibinfo{year}{2023}).

\bibitem{suli2003introduction}
\bibinfo{author}{Süli, E.} \& \bibinfo{author}{Mayers, D.~F.}
\newblock \emph{\bibinfo{title}{An introduction to numerical analysis}} (\bibinfo{publisher}{Cambridge University Press}, \bibinfo{year}{2003}).

\bibitem{trench2013elementary}
\bibinfo{author}{Trench, W.~F.}
\newblock \emph{\bibinfo{title}{Elementary differential equations with boundary value problems}} (\bibinfo{publisher}{Brooks Cole Thomson Learning}, \bibinfo{year}{2013}).

\end{thebibliography}

\clearpage

\section*{Supplementary}



\section*{Proof of the Theorem}
\begin{proof}
We denote $E_n := \hat{\mathbf{u}}_{kn} - \mathbf{u}_{kn}$. Then by triangle inequality and Proposition~{\color{blue}0.1} in the main text, we have
\begin{align}
    |\hat{\mathbf{u}}_{kp}-\mathbf{u}(T)|\leq |\mathbf{u}_{kp}-\mathbf{u}(T)|+|\hat{\mathbf{u}}_{kp}-\mathbf{u}_{kp}|=E_p+\frac{M\exp(2TL)}{2L}\Delta t.
    \label{eqn:theoryerror}
\end{align}
Next we estimate the error $E_p$. We have 
\begin{align*}
    \hat{\mathbf{u}}_{k(n+1)}-\mathbf{u}_{k(n+1)}&=\hat{\mathbf{u}}_{kn} + \mathbf{f}(\hat{\mathbf{u}}_{kn}) (k\Delta t) + \text{NeurVec}(\hat{\mathbf{u}}_{kn};\bm{\theta})-\mathbf{u}_{k(n+1)}
    \\&=\hat{\mathbf{u}}_{kn}-\mathbf{u}_{kn}+\big(\mathbf{f}(\hat{\mathbf{u}}_{kn})-\mathbf{f}(\mathbf{u}_{kn})\big) (k\Delta t) + \text{NeurVec}(\hat{\mathbf{u}}_{kn};\bm{\theta})-\text{NeurVec}(\mathbf{u}_{kn};\bm{\theta})-(k\Delta t)V_n.
\end{align*}
Then using assumption (1), 
\begin{align*}
    |\hat{\mathbf{u}}_{k(n+1)}-\mathbf{u}_{k(n+1)}|&\leq|\hat{\mathbf{u}}_{kn}-\mathbf{u}_{kn}|+ L|\hat{\mathbf{u}}_{kn}-\mathbf{u}_{kn}|(k\Delta t) + k\Delta tL_{NV}|\hat{\mathbf{u}}_{kn}-\mathbf{u}_{kn}|+(k\Delta t)|V_n|
    \\&=(1+k\Delta tL+k\Delta t L_{NV})|\hat{\mathbf{u}}_{kn}-\mathbf{u}_{kn}|+(k\Delta t)|V_n|.
\end{align*}
we denote the constant $(1+k\Delta tL+k\Delta t L_{NV})$ as $w$. We rewrite the above inequality as $|E_{n+1}|\leq w |E_{n}|+(k\Delta t)|V_n|$. Then
\begin{align*}
    |E_{n+1}|&\leq w |E_{n}|+(k\Delta t)|V_n|
    \\&\leq w(w |E_{n-1}|+(k\Delta t)|V_{n-1}|)+(k\Delta t)|V_n|=w^2 |E_{n-1}|+w(k\Delta t)|V_{n-1}|+(k\Delta t)|V_n|
    \\&\leq \cdots
    \\&\leq w^{n+1} |E_0|+(k\Delta t)\sum_{i=0}^n w^i |V_{n-i}|=(k\Delta t)\sum_{i=0}^n w^i |V_{n-i}|,
\end{align*}
where $E_0=0$ as $E_0=\hat{\mathbf{u}}_0-\mathbf{u}_0=\mathbf{c}_0-\mathbf{c}_0=0$. By the Cauchy inequality,
\begin{align*}
    |E_{p}|\leq (k\Delta t)(\sum_{i=0}^{p-1} w^{2i})^{\frac{1}{2}}(\sum_{i=0}^{p-1} |V_{p-1-i}|^2)^{\frac{1}{2}} = (k\Delta t)(\frac{w^{2p}-1}{w^2-1})^{\frac{1}{2}}(p\text{LS})^{\frac{1}{2}}.
\end{align*}
Note that $\frac{w^{2p}-1}{w^2-1}\leq \frac{(1+k\Delta tL+k\Delta t L_{NV})^{2p}}{k\Delta tL+k\Delta t L_{NV}}\leq \frac{\exp(2pk\Delta t(L+L_{NV}))}{k\Delta tL+k\Delta t L_{NV}}=\frac{\exp(2T(L+L_{NV}))}{k\Delta t(L+L_{NV})}.$
We obtain the bound by 
\begin{align*}
    |E_{p}|\leq (k\Delta t)\frac{\exp(T(L+L_{NV}))}{\sqrt{k\Delta t(L+L_{NV})}}(p\text{LS})^{\frac{1}{2}} = \frac{\sqrt{T}\exp(T(L+L_{NV}))}{\sqrt{L+L_{NV}}}(\text{LS})^{\frac{1}{2}}.
\end{align*}
Combining \eqref{eqn:theoryerror}, we  end our proof. 
\end{proof}

\newpage
\section*{Details of the acceleration.}
\subsection*{Runtime}
We presented the normalized runtime results of different configurations in the main text. Here we provide details about the implementation and the exact runtime values. 

We divide the testing dataset of each problem (Table {\color{blue}3} in the main text) equally into 70 batches. Then, we simulate each batch sequentially on a single GeForce RTX 3080 GPU and record their inference time at each run. In order to mitigate the GPU from overheating, a 10-second pause is executed between every two runs. The mean clock time and its standard derivation (std) are reported in Table~{\color{blue}S1}.

\begin{table}[htbp]
  \centering
    \begin{tabular}{lccrr}
    \toprule
    \multicolumn{1}{c}{Problems} & Method & Step size & \multicolumn{1}{c}{Time-mean (sec.)} & \multicolumn{1}{c}{Time-std} \\
    \midrule
    Spring-chain & Euler & 2e-1  & 0.099 & 0.002 \\
    Spring-chain & Euler & 1e-3  & 18.510 & 0.361 \\
    Spring-chain & Euler+NeurVec & 2e-1  & 0.124 & 0.011 \\
    \midrule
    Spring-chain & Improved Euler & 2e-1  & 0.172 & 0.002 \\
    Spring-chain & Improved Euler & 1e-3  & 38.891 & 2.831 \\
    Spring-chain & Improved Euler+NeurVec & 2e-1  & 0.237 & 0.030 \\
    \midrule
    Spring-chain & RK3   & 2e-1  & 0.260 & 0.002 \\
    Spring-chain & RK3   & 1e-3  & 57.921 & 0.416 \\
    Spring-chain & RK3+NeurVec & 2e-1  & 0.348 & 0.028 \\
    \midrule
    Spring-chain & RK4   & 2e-1  & 0.359 & 0.002 \\
    Spring-chain & RK4   & 1e-3  & 73.012 & 2.261 \\
    Spring-chain & RK4+NeurVec & 2e-1  & 0.463 & 0.040 \\
    \midrule
    Hénon-Heiles & RK4   & 5e-1  & 0.009 & 0.001 \\
    Hénon-Heiles & RK4   & 1e-3  & 5.984 & 0.097 \\
    Hénon-Heiles & RK4+NeurVec & 5e-1  & 0.015 & 0.001 \\
    \midrule
    2-link pendulum & RK4   & 1e-1  & 0.269 & 0.004 \\
    2-link pendulum & RK4   & 1e-3  & 27.801 & 0.280 \\
    2-link pendulum & RK4+NeurVec & 1e-1  & 0.441 & 0.061 \\
    \midrule
    Elastic pendulum & RK4   & 1e-1  & 0.056 & 0.001 \\
    Elastic pendulum & RK4   & 1e-3  & 7.011 & 0.094 \\
    Elastic pendulum & RK4+NeurVec & 1e-1  & 0.100 & 0.059 \\
    \bottomrule
    \end{tabular}%
  \caption*{\textbf{Table S1:} \label{tab:acce_detail} Comparison of mean runtime and its standard deviation over 70 batches of simulation. }
  \label{tab:runtime}%
\end{table}%
\newpage
\subsection*{Statistical test}

We statistically validate the acceleration performance of NeurVec. We collected the runtime results of solver with fine step size and NeurVec with coarse step size as in Table~{\color{blue}S1}; each of them contains 70 samples. In the t-test, the null hypothesis is that the runtimes of solver with fine step size and NeurVec with coarse step size are identical. We use $P_1$ to denote the $P$-value of the two-sided t-test and $P_2$ to denote the $P$-value of the Welch’s t-test. These two t-tests have the same statistics but for different situations. We use the two-sided t-test if the the pair of the runtimes have the same variance. Otherwise, we use Welch’s t-test. In Table~{\color{blue}S2}, all the $P_1$ and $P_2$ are much smaller than $1e-3$, indicating that the null hypothesis is rejected and the pair of runtime samples are different. In other words, NeurVec can accelerate the simulation with statistical significance.

\begin{table}[htbp]
  \centering
    \begin{tabular}{llllrrr}
    \toprule
    \multicolumn{1}{c}{Problem} & \multicolumn{1}{c}{Method} & \multicolumn{1}{c}{Setting1} & \multicolumn{1}{c}{Setting2} & \multicolumn{1}{c}{Statistics} & \multicolumn{1}{c}{$P_1$ ( $P_1\ll$1e-3?)} & \multicolumn{1}{c}{$P_2$ ($P_2\ll$1e-3?)} \\
    \midrule
    Spring-chain & Euler & 1e-3  & 2e-1  &  461.97     &5.59e-222 \checked       &3.55e-122 \checked \\
    Spring-chain & Euler & 1e-3  & 2e-1+NeurVec &  461.07     & 7.30e-222 \checked      & 2.44e-122 \checked \\
    Spring-chain & Imp-Euler & 1e-3  & 2e-1  &  125.34     & 4.87e-144 \checked      & 3.88e-83 \checked \\
    Spring-chain & Imp-Euler & 1e-3  & 2e-1+NeurVec & 125.11      & 6.24e-144 \checked      & 4.22e-83 \checked \\
    Spring-chain & RK3   & 1e-3  & 2e-1  & 1252.16      & 1.01e-281 \checked      & 4.73e-152 \checked \\
    Spring-chain & RK3   & 1e-3  & 2e-1+NeurVec &  1246.92     & 1.80e-281 \checked       & 2.41e-153 \checked \\
    Spring-chain & RK4   & 1e-3  & 2e-1  & 291.81      & 1.78e-194 \checked   & 2.08e-108 \checked \\
    Spring-chain & RK4   & 1e-3  & 2e-1+NeurVec & 291.32      & 2.25e-194 \checked      &  2.00e-108 \checked\\
    \midrule
    Hénon-Heiles & RK4   & 1e-3  & 5e-1  &  557.53     & 3.06e-233 \checked       & 7.91e-128 \checked \\
    Hénon-Heiles & RK4   & 1e-3  & 5e-1+NeurVec & 556.97      & 3.51e-233 \checked      & 8.48e-128 \checked \\
    \midrule
    2-pendulum & RK4   & 1e-3  & 1e-1  & 888.86      & 3.46e-261 \checked      & 8.85e-142 \checked \\
    2-pendulum & RK4   & 1e-3  & 1e-1+NeurVec & 859.68      & 3.46e-259 \checked      & 2.30e-154 \checked \\
    \midrule
    Elastic pendulum & RK4   & 1e-3  & 1e-1  & 669.28      & 3.46e-244 \checked   & 2.63e-133 \checked \\
    Elastic pendulum & RK4   & 1e-3  & 1e-1+NeurVec &  550.22     & 1.88e-232 \checked      & 7.35e-208 \checked \\
    \bottomrule
    \end{tabular}%
  \caption*{\textbf{Table S2:}\label{tab:stat} The statistical test for different pair run time results. $P_1$ is the p-value of the two-sided t-test and $P_2$ is the p-value of the Welch’s t-test. }
\end{table}%

\newpage
\section*{Summary of the initialization of different systems. }

\begin{table}[htbp]
  \centering

    \begin{tabular}{|c|ccccc|c|cccccc|}
    \toprule
    \multirow{4}[2]{*}{m} & 0.900 & 0.938 & 0.925 & 0.787 & 0.667 & \multirow{4}[2]{*}{k} & 3.900 & 3.508 & 5.651 & 5.533 & 3.664 &  \\
          & 1.348 & 0.776 & 0.692 & 0.941 & 0.538 &       & 4.373 & 2.555 & 5.239 & 6.024 & 6.942 &  \\
          & 1.215 & 0.821 & 1.121 & 0.875 & 1.456 &       & 5.073 & 3.941 & 4.505 & 4.744 & 3.805 &  \\
          & 1.111 & 1.125 & 1.431 & 0.663 & 1.222 &       & 4.848 & 3.477 & 3.405 & 2.499 & 4.735 & 4.891 \\
    \bottomrule
    \end{tabular}%
    
  \caption*{\textbf{Table S3:} \label{tab:data2}The values of $m$ and $k$ in spring-chain systems: $m$ and $k$ are obtained by random and independent sampling.}
\end{table}%

\begin{table}[htbp]
  \centering
  \resizebox{0.66\hsize}{!}{%
    \begin{tabular}{lcccc}
    \toprule
    \multicolumn{1}{c}{\textbf{Task}} & \textbf{Variable} & \textbf{Dim} & \textbf{Type} & \textbf{Range} \\
    \midrule
    Spring-chain & $\mathbf{p}$  & 20    & Uniform random & $[-2.5,2.5]^{20}$\\
    Spring-chain & $\mathbf{q}$  & 20    & Uniform random & $[-2.5,2.5]^{20}$ \\
    \midrule
    Hénon--Heiles & $q_x$    & 1     & Uniform random & [-1,1] \\
    Hénon-Heiles & $q_y$    & 1     & Uniform random & [-0.5,1] \\
    Hénon--Heiles & $p_x$    & 1     & Uniform random & [-1,1] \\
    Hénon--Heiles & $p_y$    & 1     & Uniform random & [-1,1] \\
    \midrule
    Elastic pendulum & $\theta$ & 1     & Uniform random & $[0,\pi/8]$ \\
    Elastic pendulum & $r$     & 1     & Constant & 10 \\
    Elastic pendulum & $\dot{\theta}$ & 1     & Constant & 0 \\
    Elastic pendulum & $\dot{r}$  & 1     & Constant & 0 \\
    Elastic pendulum & $l_0$    & 1     & Constant & 10 \\
    Elastic pendulum & $g$     & 1     & Constant & 9.8 \\
    Elastic pendulum & $k$    & 1     & Constant &  40\\
    Elastic pendulum & $m$     & 1     & Constant &  1\\
    \midrule
    2-pendulum & $\mathbf{\theta}$ & 2     & Uniform random & $[0,\pi/8]^2$ \\
    2-pendulum & $\mathbf{\dot{\theta}}$ & 2     & Constant & 0 \\
    1-pendulum & $\theta$ & 1     & Uniform random & $[0,\pi/2]$ \\
    1-pendulum & $\dot{\theta}$ & 1     & Uniform random & [0,0.5] \\
    1\&2-pendulum & $m$     & 1     & Constant & 1 \\
    1\&2-pendulum & $g$     & 1     & Constant & 9.8 \\
    \bottomrule
    \end{tabular}%
    }
  \caption*{\textbf{Table S4:} \label{tab:init}Initial state of different systems. ``Uniform random'' means that the variables are sampled with uniform distribution of given range. ``Constant'' means the variable is initialized as a constant. In the Hénon--Heiles system, after the initialization, the data that do not satisfy the energy  $\mathcal{H}\left(q_x,q_y,p_x,p_y\right) \in [\frac{1}{12},\frac{1}{6}]$ will be removed. \rv{The value of $\theta$ significantly influences the difficulty of solving the pendulum system, for a more detailed discussion, please refer to Fig. {\color{blue}S6}.}}
\end{table}%

\newpage
\section*{Simulations on Hénon--Heiles system.}
Fig.~{\color{blue}S1} displays more examples of simulations on Hénon--Heiles system, an implement to Fig.~{\color{blue}3}e in the main text.

\begin{figure}[H]
    \centering
    \includegraphics[width=0.94\linewidth]{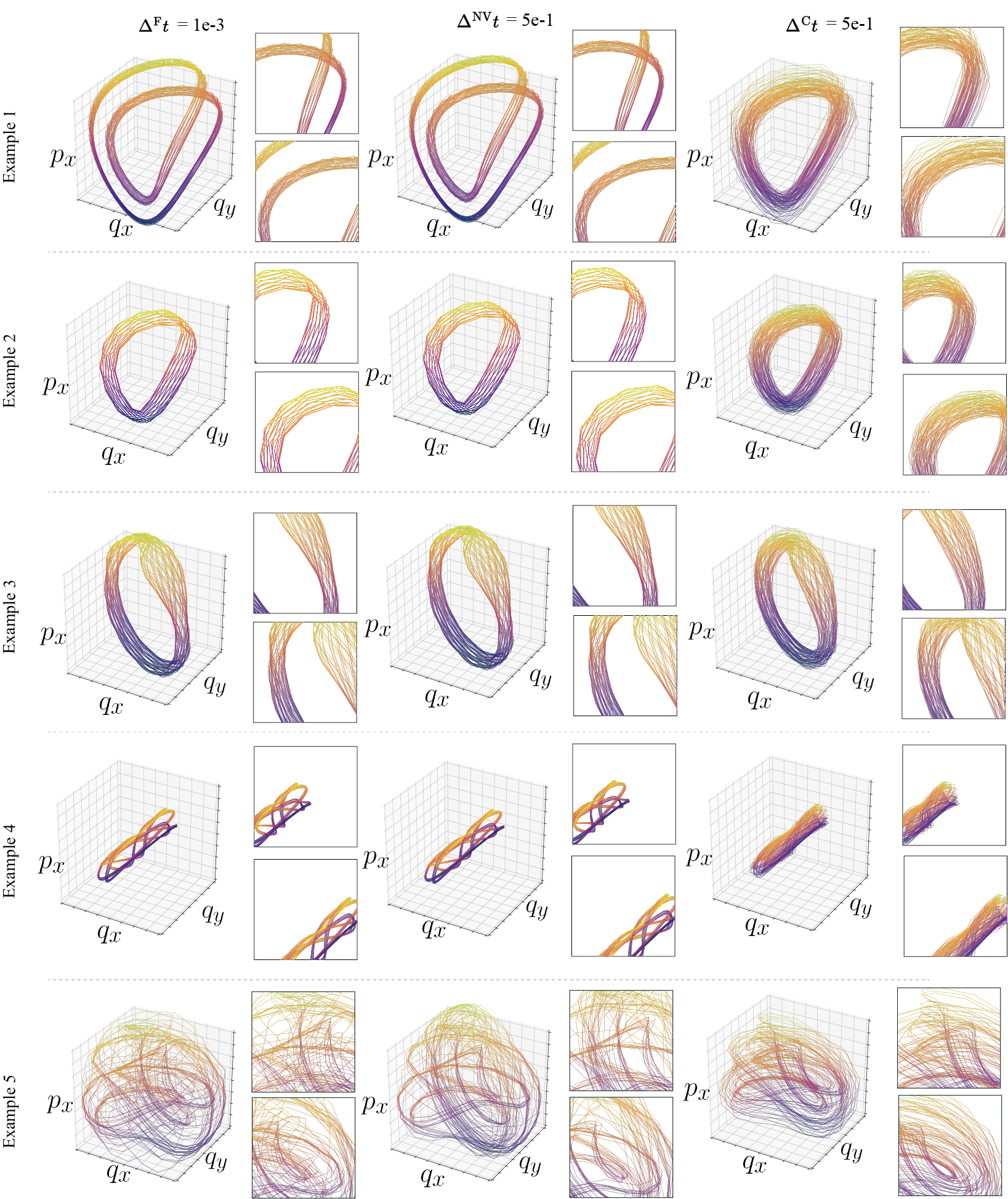}
    \caption*{\textbf{Figure S1:} \textbf{More performance comparison on the Hénon--Heiles system.}  We provide five additional examples of trajectories projected on the coordinates $(q_x,q_y,p_x)$. NeurVec ($\Delta^{\rm NV}t = 5e-1$) produces the
most  orbits similar to $\Delta^{\rm F}t = 1e-3$. }
    \label{fig:moreexample}
\end{figure}

\section*{Time series histogram.}
Fig.~{\color{blue}S2} displays the time series histogram of $\theta$ and $\dot{\theta}$ in the elastic pendulum, i.e., Eq.~({\color{blue}10}) in the main text, and $q_x$ and $p_x$ in the Hénon--Heiles system, i.e., Eq.~({\color{blue}8}) in the main text. The statistical difference of $\theta$ and $\dot{\theta}$ among fine step size, coarse step size and NeurVec with coarse step size is not large. Yet that of $q_x$ and $p_x$ is large. 
\begin{figure*}[h]
    \centering
    \includegraphics[width=\linewidth]{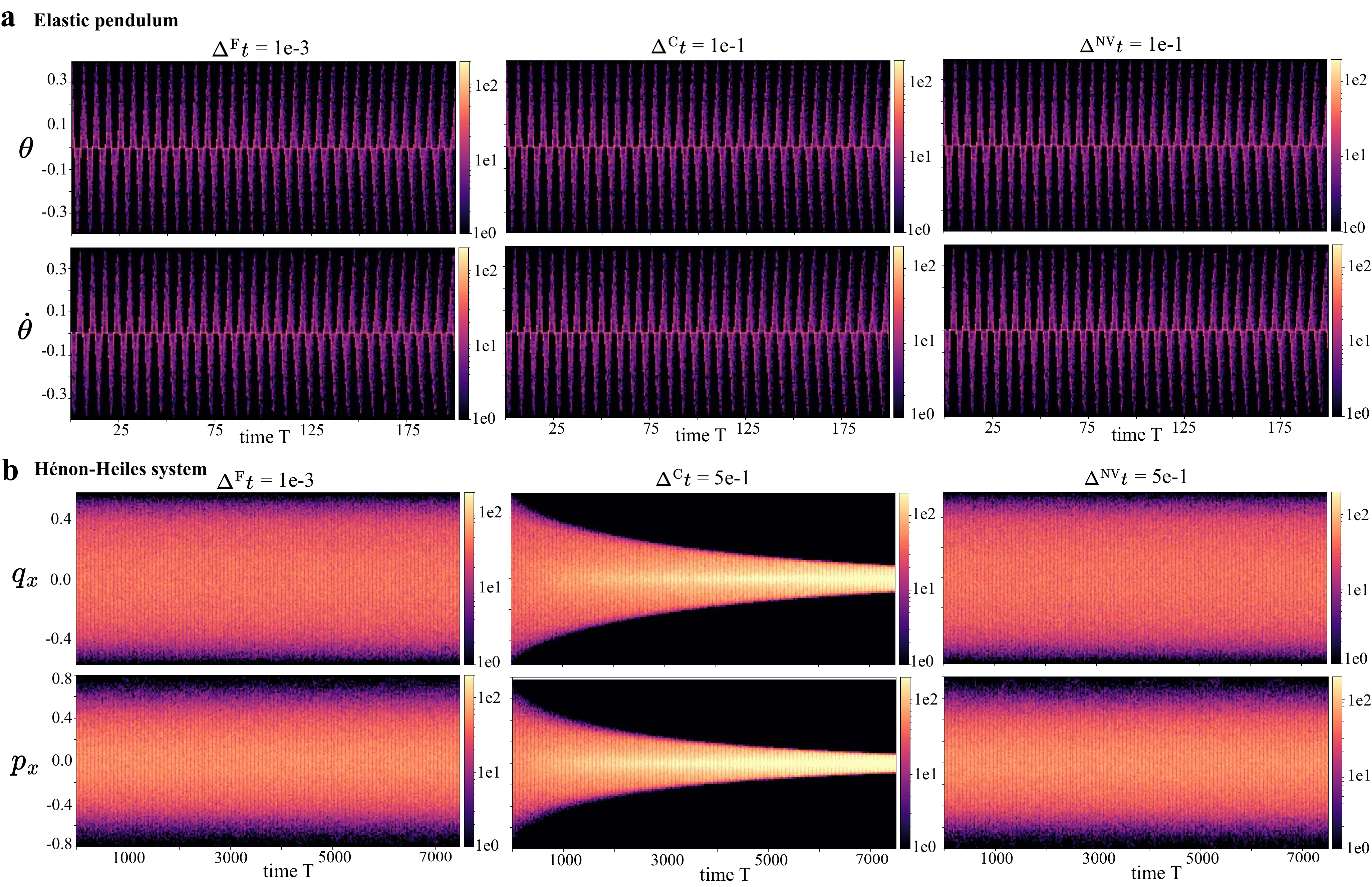}
    \caption*{\textbf{Figure S2:} \textbf{Additional experiments for time series histogram.} We visualize the time series histogram of the test set for variables in the elastic pendulum and Hénon--Heiles system. The color represents the number count (the lighter color and the larger frequency).
    (\textbf{a}), $\theta$ and $\dot{\theta}$ in the elastic pendulum, i.e., Eq.~({\color{blue}10}) in the main text. Unlike the results about $r$ and $\dot{r}$ in Fig.~{\color{blue}5}a of the main text, the $\theta$ and $\dot{\theta}$ generated by different step sizes have similar trends, although there are minor differences among them. However, in   
    (\textbf{b}) $q_x$ and $p_x$ in the Hénon--Heiles system, i.e., Eq.~({\color{blue}8}) in the main text, there is a consistent observation for the solutions under different step sizes with Fig.~{\color{blue}5}b in the main text.  }
    \label{fig:66}
\end{figure*}

\newpage
\section*{\rv{The distribution of training data for 1-link pendulum.}}

\rv{In Fig.~{\color{blue}S3}, we present the training data for the analysis in Figure 6 in the main text, where the background illustrates the difference between the leading error term and NeurVec. The relatively significant differences near the boundary might be attributed to the limited availability of data that encompasses those boundary regions. Furthermore, in regions with limited training data, such as when $\dot{\theta} = -1.0$ and $\theta = 1.5$, the high accuracy demonstrated by our NeurVec also indicates to some extent the generalization capability.}
\begin{figure}[htbp]
    \centering
    \includegraphics[width=0.8\linewidth]{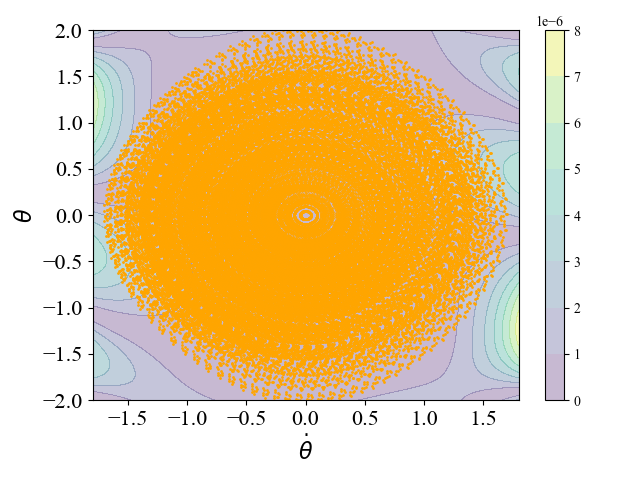}
    \caption*{\textbf{Figure S3:} \rv{The training data (orange dots) used in 1-link pendulum. The background represents the difference between the leading error term and NeurVec.} }
    \label{fig:trainingdata}
\end{figure}


\newpage
\section*{Application of NeurVec on the Kuramoto-Sivashinsky equation.}
\rv{We intend to showcase NeurVec's potential in handling more intricate scenarios by presenting its numerical results on the Kuramoto-Sivashinsky equation (KSE), thus demonstrating its potential to more challenging systems. 

We consider KSE on a domain with periodic boundary conditions. The equation is given by:
\begin{align*}
\frac{du}{dt} = -uu_x - u_{xx} - u_{xxxx}, \ \ x\in [0, L]. 
\end{align*}
Here, $L$ represents the domain length, and we choose $L = \frac{2\pi}{\sqrt{0.085}}$. To solve this equation numerically, we use the exponential time-differencing fourth-order Runge-Kutta method (ETDRK4) as a forward scheme $S$. The spatial variable $x$ is discretized with a resolution of 48 uniformly spaced points on the interval $[0, L]$.

Our training dataset consists of a single trajectory with an initial state given by $u(x,0)=\cos((\pi/L)x)$. This trajectory is generated with a time step size of $\Delta^{\rm F} t = 2 \times 10^{-2}$, and the model time reaches $T = 200$k. For testing, we have reference simulations with a smaller time step size of $5 \times 10^{-3}$.

NeurVec is trained using the simulations with $\Delta^{\rm F} t = 2 \times 10^{-2}$ to learn the error correction of ETDRK4 with a coarse time step size $\Delta^{\rm NV} t = 1$. To assess accuracy, we use the mean square error (MSE) between the reference solution and the simulated solution of $\Delta^{\rm NV} t = 1$, averaged over 20 different initializations.

Fig. {\color{blue}S4}a presents the MSE, where the solid curves and the shaded areas represents the mean and one standard deviation, respectively, calculated from 20 simulation runs. We can observe that NeurVec manages to maintain short-term accuracy, closely matching the reference solution, up until the model time reaches $T=75$. However, beyond this point, NeurVec starts to deviate significantly from the reference solution, yet still maintains smaller error than the simulations of $\Delta^{\rm C} t = 1$. Fig. {\color{blue}S4}b shows two examples of the visualized solution for a short period of time. Even though the highly chaotic nature of KSE, NeurVec can still maintain the short-term accuracy as the reference one. 

\begin{figure}[htbp]
    \centering
    \includegraphics[width=0.8\linewidth]{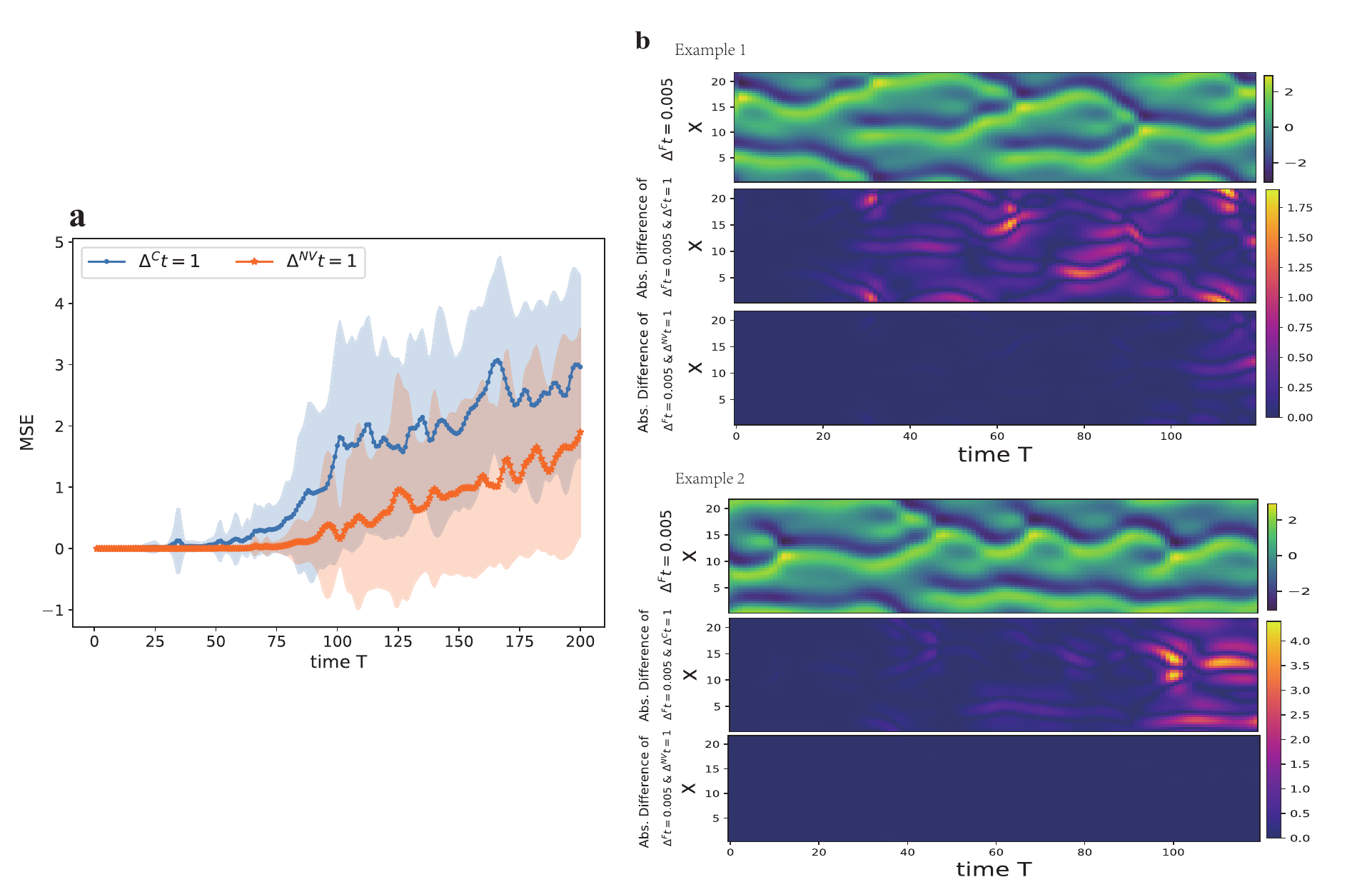}

    \caption*{\textbf{Figure S4:} \rv{\textbf{a}, The mean square error (MSE) between the reference solution and the numerical solutions with different configurations (step size $\Delta^{\rm C}t = 1$ and NeurVec  with $\Delta^{\rm NV}t = 1$) for KSE. \textbf{b}, we provide two examples of different initializations and plot the absolute differences. } }

    \label{fig:kse}
\end{figure}}

\newpage
\section*{Impact of neural network structures on $\epsilon$.}
\rv{The $\epsilon$ is used to denote the runtime ratio of NeurVec to the scheme $S$. This section delves into how the neural network structures influence the value of $\epsilon$. 

Taking 2-link pendulum system with the Fourth-Order Runge-Kutta method as an example, in Fig. {\color{blue}S5}, we present the values of $\epsilon$ in neural networks under different hidden dimensions, numbers of layers, and activation functions on one A100 GPU. We can observe that the choice of neural network structure significantly impacts the acceleration effect of the neural network on the dynamical system. Specifically,

(1) For the hidden dimension, we use a default setting of 1024. From Fig. {\color{blue}S5}, we can find that larger hidden dimensions lead to slower inference speed of the neural network.

(2) Regarding the number of layers, we notice that this variable has a greater influence on inference speed compared to the hidden dimension. When the number of layers reaches 4, $\epsilon$ is more than twice that of our default setting (with 2 layers). Therefore, selecting the number of layers requires more careful consideration.

(3) As for the activation function, different choices also introduce variations in the model's inference efficiency. Despite the higher inference cost of the ``Rational'' activation function, it equips the model with strong non-linear fitting capabilities to tackle the complex representation learning of dynamical systems. 
When the neural networks operate without an activation function (i.e., ``Identity''), or when using ReLU, they experience a performance drop of approximately 1$\sim$2 orders of magnitude in terms of MSE on the test set, although they exhibit slightly faster inference speeds.

For the dynamical systems considered in this paper, we find that selecting a hidden dimension of 1024, 2 layers, and the ``Rational'' activation function provides a sufficiently favorable acceleration effect for solving these systems. When addressing more intricate scenarios in the future, such as complex equations or higher energy levels, careful consideration of the model's structure will be necessary to enable NeurVec to deliver optimal acceleration performance.

}

\begin{figure}[htbp]
    \centering
    \includegraphics[width=0.9\linewidth]{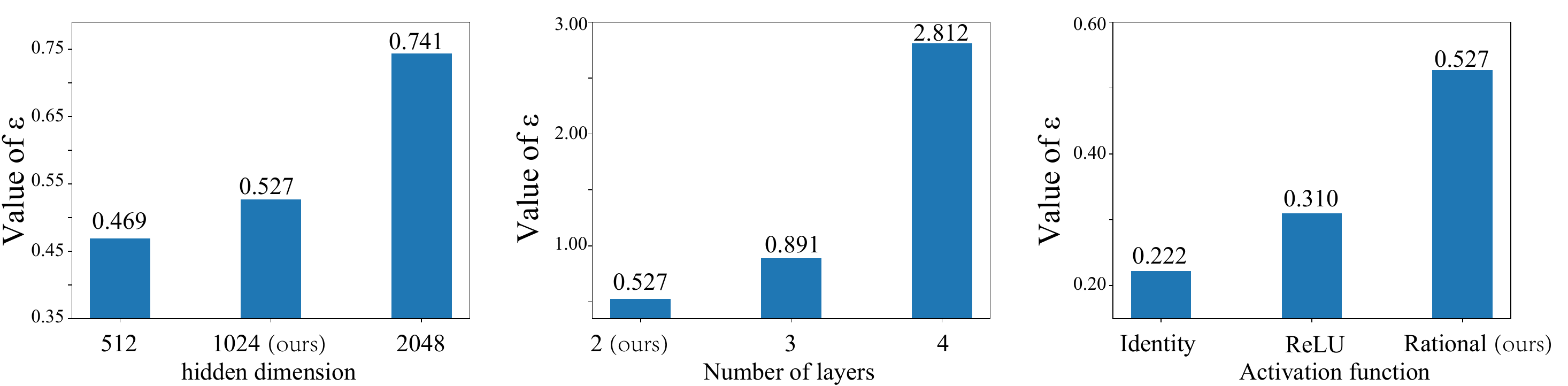}
     \caption*{\textbf{Figure S5:} \rv{The value $\epsilon$ with different network structures.}}
    \label{fig:epsilon}
\end{figure}

\newpage
\section*{Exploring NeurVec's performance on trajectories with varying energy levels}
\rv{
In order to explore NeurVec's ability to handle trajectories with varying energy levels, 
we kept all variables fixed in initial conditions except for $\theta$ in Elastic Pendulum and the 2-link Pendulum, and varied the range of $\theta$ by changing its sampling interval to investigate the boundaries of NeurVec's capabilities. The sampling interval is $[0,\theta_{\max}]$. Generally, as the upper bound $\theta_{\max}$ of the initial $\theta$ increases, trajectories with higher energy levels will be generated, posing a greater challenge for the neural network's learning process. In our paper, $\theta_{\max}$ was set to $\frac{\pi}{8}$. Specifically, for the Elastic Pendulum, $\theta$ was uniformly sampled from $[0,\frac{\pi}{8}]$ in the initial conditions. For the 2-link Pendulum, $\theta$ was sampled from a two-dimensional uniform distribution $[0,\frac{\pi}{8}]^2$.

Fig. {\color{blue}S6} visually represents NeurVec's training Mean Squared Error (MSE) at the 300th epoch for different values of $\theta_{\max}$. The trajectories with higher energy levels (i.e., larger $\theta_{\max}$) are more challenging to learn due to their increased chaotic behavior. From Fig. {\color{blue}S6}, we can observe that currently, NeurVec is capable of tackling $\theta_{\max}$ of approximately less than $\frac{\pi}{4}$.} 
\begin{figure}[htbp]
    \centering
    \includegraphics[width=0.6\linewidth]{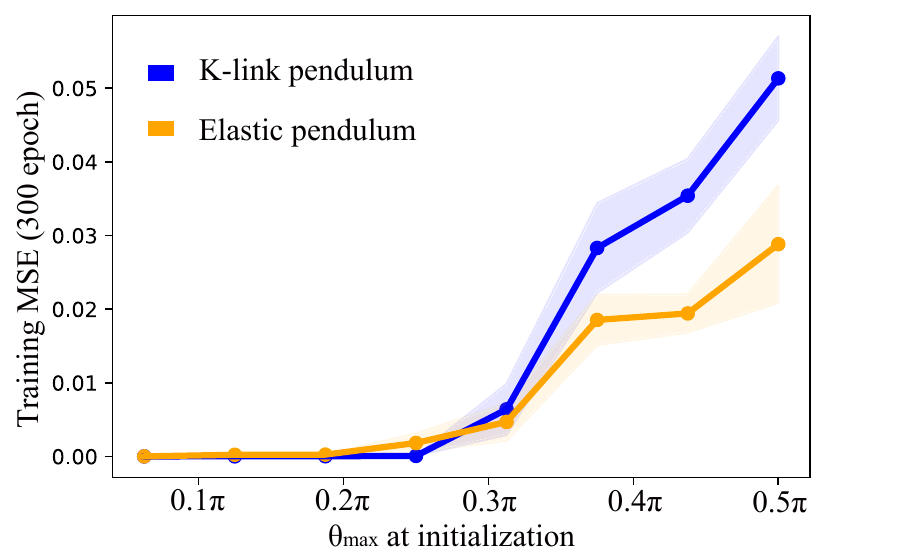}
    \caption*{\textbf{Figure S6:} \rv{The training MSE on different $\theta_{\max}$.}}
    \label{fig:energybound}
\end{figure}

\end{document}